\documentclass[
 cha,
 aip,
 amsmath,amssymb,
 reprint,%
]{revtex4-1}

\usepackage{graphicx}
\usepackage{dcolumn}
\usepackage{bm}

\usepackage[utf8]{inputenc}
\usepackage[T1]{fontenc}
\usepackage{mathptmx}
\usepackage{etoolbox}
\usepackage{subcaption}

\makeatletter
\def\@email#1#2{%
 \endgroup
 \patchcmd{\titleblock@produce}
  {\frontmatter@RRAPformat}
  {\frontmatter@RRAPformat{\produce@RRAP{*#1\href{mailto:#2}{#2}}}\frontmatter@RRAPformat}
  {}{}
}%
\makeatother

\usepackage{bm}
\usepackage{tikz}

\newcommand{\R}{\mathbb{R}}

\newcommand{\eps}{\varepsilon}

\begin{document}


\title[Motion in Aubry's galaxy]{Motion in Aubry's galaxy}

\author{M.~Burri}
\email{burri.megha@gmail.com}
\affiliation{Physics department, University of Warwick, Coventry CV4 7AL, U.K.}

\author{R.~S.~MacKay}%
\email{R.S.MacKay@warwick.ac.uk}
\affiliation{Mathematics Institute \& Centre for Complexity Science, University of Warwick, Coventry CV4 7AL, U.K.  Corresponding author}


\date{\today}

\begin{abstract}
The dynamics of a test particle in the field of a model galaxy proposed by Serge Aubry is studied by a combination of theory and numerical computation.  Regimes of near-integrable behaviour and almost perfect chaos are found.  A proposed explanation for the latter is sketched.  Thoughts are presented on the implications of the analysis for galaxies.
\end{abstract}

\maketitle

\begin{quotation}
Aubry proposed a model for motion  in a galaxy consisting of the usual ``electrostatic'' monopole of a central mass and an additional ``magnetic'' dipole corresponding to rotating mass near the centre in a gravitomagnetic formulation.  We analyse the resulting motion by a combination of theory and numerics.   We find a spectrum of behaviour from near-integrable to near-perfect chaos.
\end{quotation}

\section{\label{sec:intro}Introduction}

Serge Aubry has been creative in many areas of physics and mathematics.  His latest contribution \cite{A} is to note that Newtonian gravity is not Lorentz-invariant and therefore there should also be a ``magnetic'' contribution to gravity, whereby moving matter exerts a force of magnetic type on moving matter.  As he has commented, this idea was already proposed by Heaviside and has been developed by various people, notably Jefimenko \cite{J} (and see ref.~\onlinecite{BB} for a recent article).  Furthermore, it seems to be an established view in Newtonian approximation to general relativity, e.g.~ref.~\onlinecite{CT}. Yet Aubry brings incisive views to the story. 

One particular consequence Aubry proposed is that a simple model for a galaxy could consist of a central mass of which at least part is rotating, thereby creating a gravitational ``magnetic'' dipole field in addition to the standard ``electrostatic'' monopole.  Aubry was interested in non-relativistic motion in this field, with a view to explaining the rotation curves of galaxies, but given the relativistic origin of the magnetic force it seems natural to extend to relativistic motion.  

An immediate question is whether the gravitational force on a test particle should be taken proportional to its rest mass or its relativistic mass.  Following Jefimenko, we take rest mass, though Aubry has been investigating the case of relativistic mass.  For the rotation curves, probably only the non-relativistic regime is relevant, where both cases agree.

In the case of gravitational force proportional to rest mass, the resulting equations of motion for the position $\bm{r}$ and velocity $\bm{v}$ of a test particle, or momentum $\bm{p} = \gamma \bm{v}$ per unit rest mass, where 
\begin{equation}
\gamma = \sqrt{1+p^2/c^2} = 1/\sqrt{1-v^2/c^2},
\label{eq:gamma}
\end{equation} with respect to coordinate-time $t$, are
\begin{eqnarray}
\frac{d \bm{p}}{dt} &=& -GM \frac{\bm{r}}{r^3} + \frac{GD}{c^2} \bm{v} \times \left(3 z \frac{\bm{r}}{r^5} - \frac{\bm{ \hat{z}}}{r^3}\right) \label{eq:pdot}\\
\frac{d \bm{r}}{dt} &=& \bm{v}, \label{eq:rdot}
\end{eqnarray}
where $M$ is the central mass, $D$ is the dipole moment (its relativistic angular momentum in the $+z$ direction), $G$ is the gravitational constant and $c$ is the speed of light.  
We take $D\ge 0$.

An alternative description of our system is as the relativistic motion of a charged particle (with $e/m = 1$) in the fields $$\bm{E} = -GM\frac{\bm{r}}{r^3}$$ of an electrostatic monopole and $$\bm {B} = \frac{GD}{c^2} \left(3z\frac{\bm{r}}{r^5} - \frac{\bm {\hat{z}}}{r^3}\right)$$
of a magnetic dipole in Minkowski space.
The case with only magnetic dipole has been studied by many people, going back to St\"ormer in connection with charged particle motion in the earth's magnetic field.  For a recent treatment, see ref.~\onlinecite{L+}.  As we will scale $M$ to 1, the dipole-only case will correspond to the limit $D = \infty$ (with appropriate scaling of other quantities).  The problem of motion of charged particles in the earth's magnetic field, however, should really include a gravitational monopole field, even if its effect is small compared to the dipole field, so our treatment is relevant if one takes $D$ large.  On the other hand, the earth's magnetic field has significant deviations from axisymmetry, but we don't treat that here.

If preferred, one can write the equations with respect to proper time $\tau$ for the particle by using $dt = \gamma\, d\tau$:
\begin{eqnarray}
\frac{d \bm{p}}{d\tau} &=& -\gamma GM \frac{\bm{r}}{r^3} + \frac{GD}{c^2} \bm{p} \times \left(3 z \frac{\bm{r}}{r^5} - \frac{\bm{\hat{z}}}{r^3}\right) \\
\frac{d \bm{r}}{d\tau} &=& \bm{p}. 
\end{eqnarray}

Centripetal acceleration in coordinate time for circular motion in a circle of radius $R$ at speed $v$ is $\gamma v^2/R$, 
so this system has circular equatorial orbits with velocity component $v_\phi$ in the tangential direction (in this paper we use ``physical'' components for vectors, as opposed to contravariant or covariant components) satisfying 
\begin{equation}
\frac{\gamma v_\phi^2}{R} =\frac{GM}{R^2} + \frac{GDv_\phi}{c^2R^3}.
\label{eq:circeq}
\end{equation}
From the definition (\ref{eq:gamma}) of $\gamma$, squaring (\ref{eq:circeq})  transforms it to a quartic in $v_\phi$ given $R$.  It does not have clean solutions, but by considering the graphs of the two sides as functions of $v_\phi \in (-c,c)$, we see that there are precisely two solutions $v_\phi$ for each radius $R$, one positive, one negative, corresponding to orbits that are co- and counter-rotating with respect to the angular momentum of the central mass.  Alternatively, one can consider it as a quadratic in $R$ for given $v_\phi \ne 0$ and (supposing $D>0$) obtain one positive solution for $R$ if $v_\phi>0$ and two positive solutions if $v_\phi \in (v_{\min},0)$, where $v_{\min}<0$ is the solution of 
\begin{equation}
\gamma v_{\min}^3 = -\frac{GM^2c^2}{4D}
\label{eq:vmin}
\end{equation}
(remember that $\gamma$ depends on $v$).

Aubry posed the question of what the other orbits look like.  Here we answer this question by a combination of theory and numerics.

To simplify treatment, we use units for length, time and mass in which $G=1$, $c=1$, $M=1$ (instead of making $c=1$ one can make $D=1$, which would be useful in the non-relativistic regime, but we did not use that scaling).  

\begin{figure}[htbp]
    \centering
    \begin{subfigure}[b]{0.3\textwidth}
        \centering
        \includegraphics[height=2.4in]{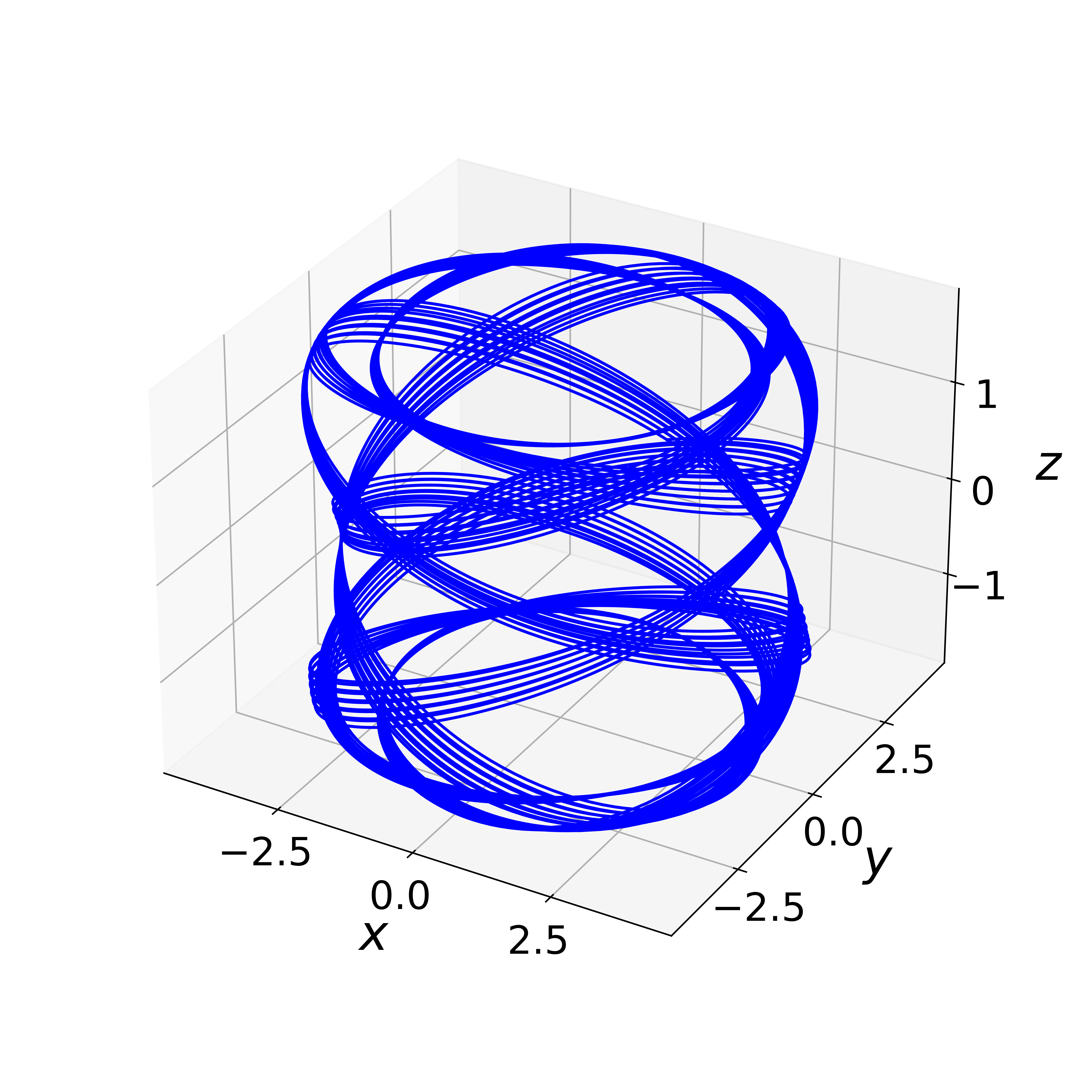}
        \caption{$(E, L_z, D)=(0.837, -0.864, 3.10)$}
    \end{subfigure}
    \hfill
    \begin{subfigure}[b]{0.3\textwidth}
        \centering
        \includegraphics[height=2.4in]{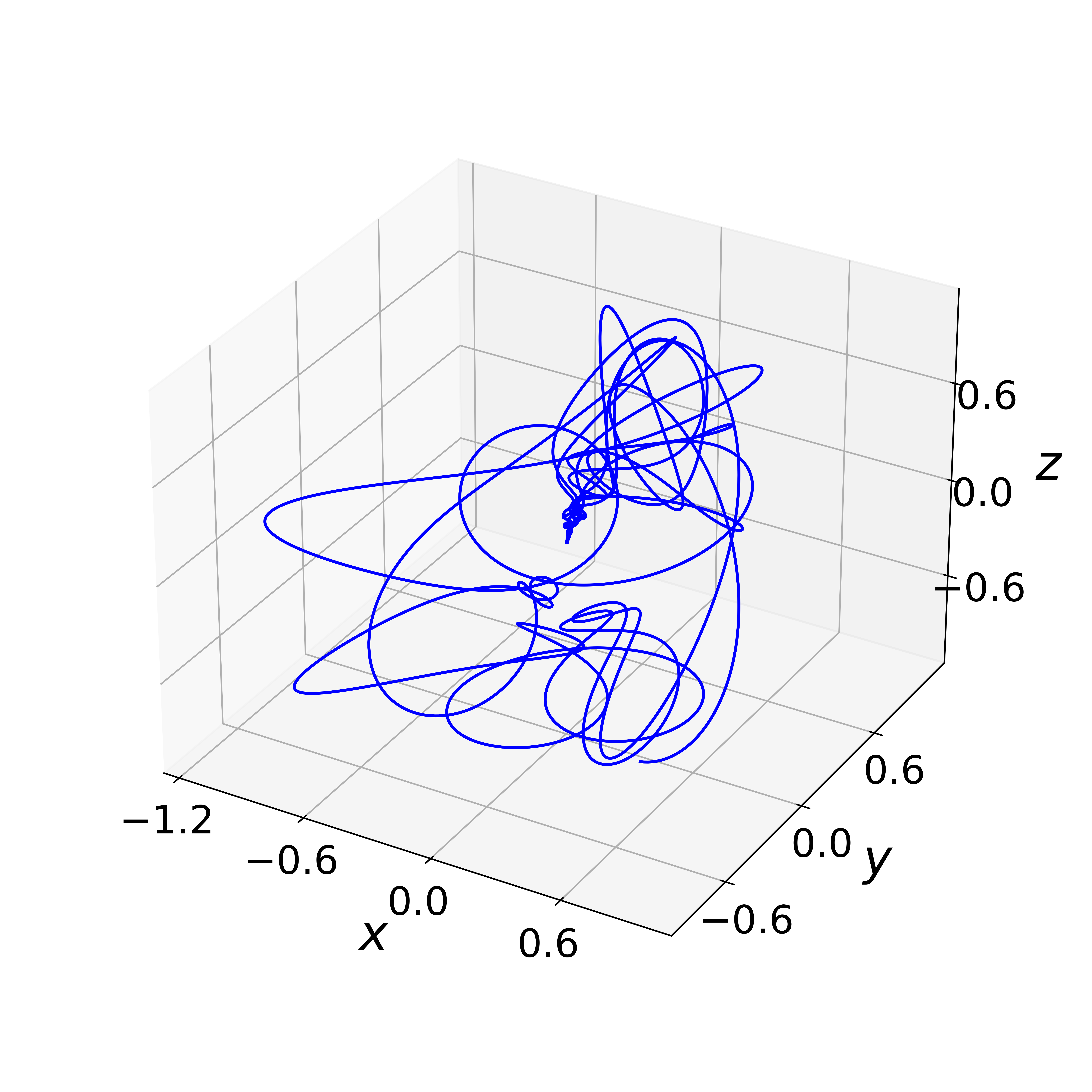}
        \caption{$(E, L_z,D)=(0.225, 0.200, 0.700)$}
    \end{subfigure}
    \hfill
    \begin{subfigure}[b]{0.3\textwidth}
        \centering
        \includegraphics[height=2.4in]{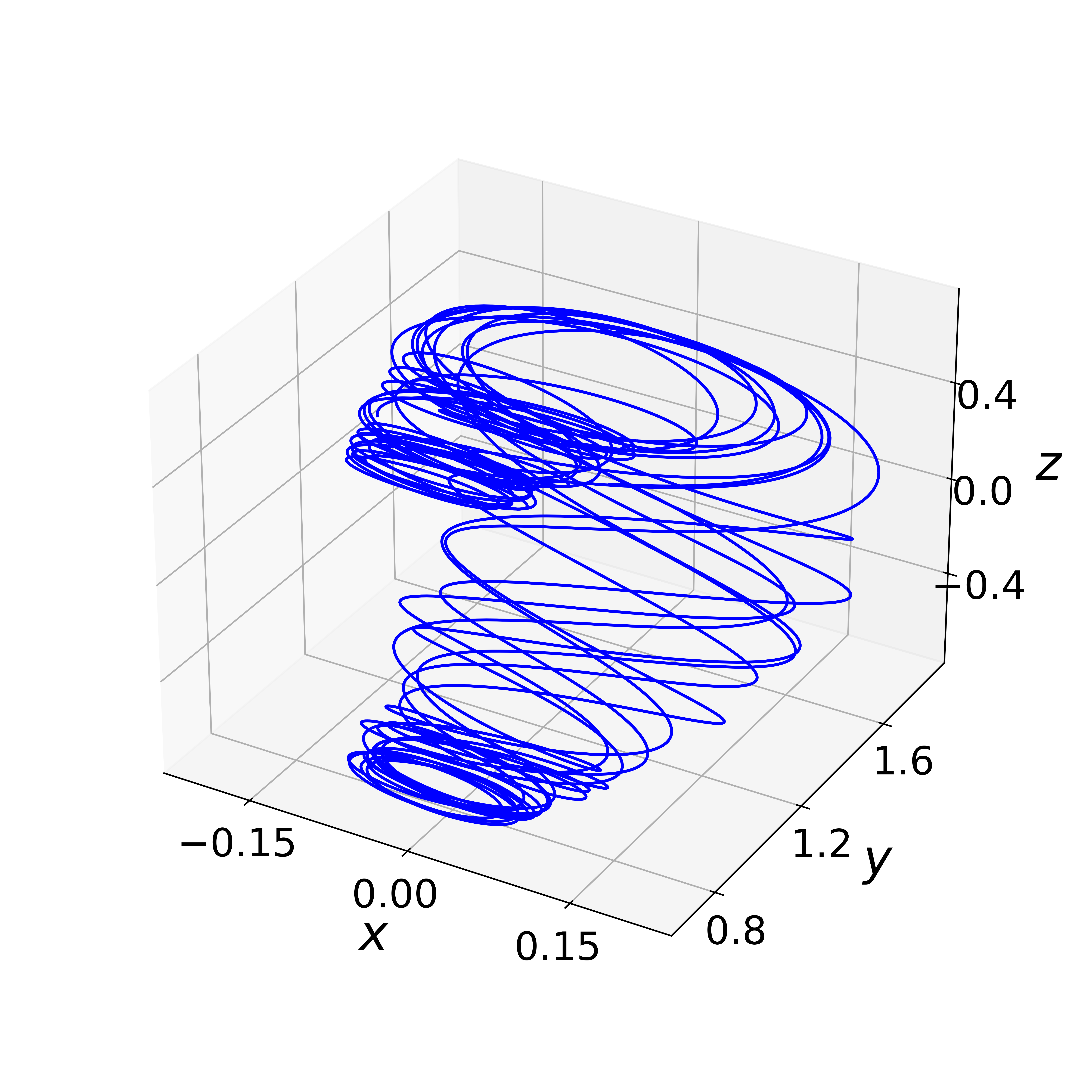}
        \caption{$(E,L_z,D)=(0.558, 5.92, 10.0)$}
    \end{subfigure}
    \caption{Sample trajectories for various parameters exhibiting a range of behaviours:~a) near integrable, b) chaotic, and c) helical.}
    \label{fig:orbits}
\end{figure}

To gain a first idea of the dynamics, we plot some trajectories in $(x,y,z)$ in Figure~\ref{fig:orbits}.
We highlight some features already:~they are often constrained to some region of space, they may be quasiperiodic or chaotic, and some exhibit helical motion when close to the centre.  We will address these features among others.

\section{Constants of the motion}

To address the constraints on the region of space for an orbit, we point out some constants of the motion.

There is an obvious constant of the motion, the energy
\begin{equation}
H(\bm{r},\bm{p}) = \gamma - \tfrac{1}{r}.
\label{eq:H}
\end{equation}
We denote its value by $E$ (there is a potential notational clash with the absolute value of the electrostatic field ${\bf E}$ but we will write $|\bm{E}|$ for the latter).
We will be interested mainly in the orbits with $H = E < 1$, the analogue of elliptic orbits in the Kepler problem, as opposed to the hyperbolic ones.  Then $r \le \frac{1}{1-E}$ so the motion is bounded by at least this condition.  But we will address $E\ge 1$ too. 

There is another constant of the motion associated with rotation symmetry about the $z$-axis, which is easiest described in cylindrical polar coordinates $(R,z,\phi)$.   It is an ``angular momentum'' constant
\begin{equation}
L_z = R (p_\phi +A_\phi),
\label{eq:Lz}
\end{equation}
where $\bm{A}$ is a vector potential for the dipole field, chosen to be independent of $\phi$, namely $\bm{A} = A_\phi \bm{\hat{\phi}}$ with
\begin{equation}
A_\phi = D\tfrac{R}{r^3}.
\end{equation}
It follows that
\begin{equation}
Rp_\phi = {L_z}-\tfrac{DR^2}{r^3}.
\label{eq:Rpphi}
\end{equation}
Note that $r^2 = R^2+z^2$.

In the limit case $D=0$, there is an additional independent constant of the motion, the square of the angular momentum vector, $L^2$, or one can equivalently take $L_x^2 + L_y^2 = (p_yz-p_zy)^2 + (p_zx-p_xz)^2$, because we already have $L_z$ constant.  This follows from spherical symmetry and can be verified directly.
One question we address is whether this third constant of the motion has a continuation to $D>0$. For geodesic motion in the Kerr metric, which can be considered the general relativistic version of Aubry's field,
 there is indeed a third constant of the motion, the ``Carter constant'' \cite{Ca}.
We find, however, that the answer for Aubry's field is no (at least from numerical simulations),

\section{Equatorial circular orbits}
\label{sec:eqcirc}

To illustrate the ideas so far, we compute the energy and ``angular momentum'' for the equatorial circular orbits.  In this section, we write $v$ for $v_\phi$.

Transforming (\ref{eq:circeq}) to scaled variables and solving the resulting quadratic in $R$, the circular orbits are given by 
\begin{equation}R = \frac{1 + \sigma \sqrt{1+4D\gamma v^3}}{2\gamma v^2},
\label{eq:eqR}
\end{equation}
with $\sigma= \pm$ for $v<0$, $+$ for $v>0$.  The formula
$$\frac{1}{R} = \frac{-1 + \sigma \sqrt{1+4D\gamma v^3}}{2Dv}$$
will also be useful.  

\begin{figure}[htbp] 
   \centering
   \includegraphics[width=2.5in]{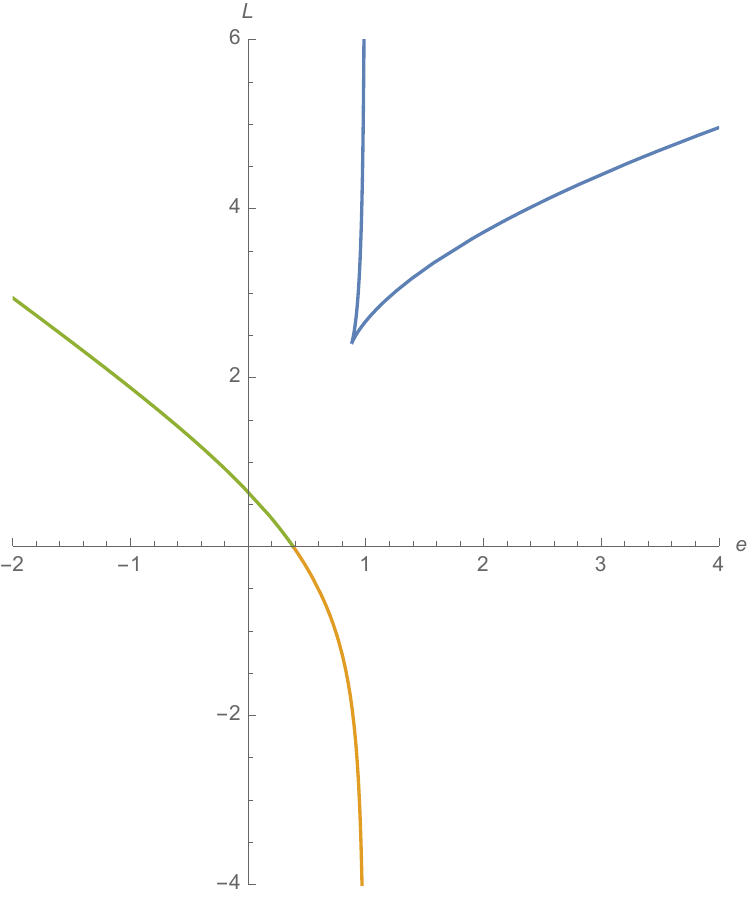} 
   \caption{The set of circular orbits for $D = 1$ projected into the plane of $(E,L_z)$.
   The blue curve is for $v_\phi=v>0$; it starts from $v=0$ at $L_z=+\infty$, $E=1$ with $R=\infty$, and as $v \to 1$ it goes to $R=0$ with $E$ and $L_z$ going to $+\infty$.  The red/green curve is for $v<0$ with the switchover at the minimum possible speed $v_{\min}$ (\ref{eq:vmin}) (which occurs at $L_z=0$); $v \to 0$ at both ends; $R$ goes from 0 at the end of the green part to $+\infty$ at the end of the red part. }
   \label{fig:circ}
\end{figure}

Its energy is
\begin{equation}
E = \gamma - \frac{1}{R} = \gamma + \frac{1}{2Dv}(1-\sigma \sqrt{1+4D\gamma v^3}),
\label{eq:eqE}
\end{equation}
and its angular momentum constant is
\begin{equation}
L_z = \gamma Rv + \frac{D}{R} = \frac{\sigma}{v} \sqrt{1+4D\gamma v^3}.
\label{eq:eqL}
\end{equation}
For a parametric plot, see Figure~\ref{fig:circ}. 

The cusp can be understood as resulting from a double root of the equation for critical points of $H$ with respect to $R$ for $z=0, p_z=0, p_R=0$ and fixed $L_z$, to be addressed in section~\ref{sec:equator}.

Note that if we think of speeds $v$ for which $\gamma v^3/c^3 \approx \frac14$ and we put back the constants $G,M,c$, then the argument of the square roots above makes a transition from the first to the second term being dominant as $D$ increases through $GM^2/c$, which is the Kerr limit for the angular momentum.  The connection, however, is probably no deeper than dimensional analysis.

\section{Hamiltonian formulation}

It is convenient to exploit a Hamiltonian formulation of the dynamics.  In addition to the {\em Hamiltonian} $H$ as in (\ref{eq:H}), which we rewrite here:
\begin{equation}
H = \gamma - \tfrac{1}{r},
\end{equation}
define the symplectic form (a {\em symplectic form} is a non-degenerate closed antisymmetric bilinear map from pairs of tangent vectors to $\R$)
\begin{equation}
\omega = \sum_i dr_i \wedge dp_i - dA^\flat
\end{equation}
in Cartesian coordinates,
where $A^\flat$ is the 1-form 
\begin{equation}
A^\flat = \tfrac{D}{r^3}(x\, dy-y\, dx).
\end{equation}

The equations of motion are  $(\dot{\bm{r}},\dot{\bm{p}}) = X(\bm{r},\bm{p})$, where $X$ is the vector field on $\R^3\times \R^3$ such that $i_X\omega = dH$.  We check it agrees with (\ref{eq:pdot}, \ref{eq:rdot}).  First evaluate
\begin{equation}
dA^\flat = 2\tfrac{D}{r^3} dx \wedge dy - 3 \tfrac{D}{r^5} (R^2 dx \wedge dy - yz\, dz \wedge dx + xz\, dz \wedge dy).
\end{equation}
Applying $i_X\omega = dH$ to $\partial_{p_i}$ yields the easy equations $\dot{r}_i = \frac{p_i}{\gamma}$.
Applying $i_X\omega = dH$ to $\partial_{r_i}$ for $i=x,y,z$, in turn we obtain
\begin{eqnarray}
\dot{p}_x &=& -\frac{1}{r^3}x + \frac{D}{r^5} (2r^2-3R^2)\frac{p_y}{\gamma} - 3 \frac{D}{r^5} yz \frac{p_z}{\gamma} \\
\dot{p}_y &=& -\frac{1}{r^3} y - \frac{D}{r^5} (2r^2-3R^2) \frac{p_x}{\gamma} + 3\frac{D}{r^5} xz \frac{p_z}{\gamma} \\
\dot{p}_z &=& -\frac{1}{r^3} z + 3\frac{D}{r^5} \frac{z}{\gamma} (yp_x-xp_y) .
\end{eqnarray}
Noting that $2r^2-3R^2 = 3z^2-r^2$, we see we have the desired equations of motion.
Thus, we have formulated the system as a Hamiltonian system of 3 degrees of freedom (DoF).

Note that the equations of motion are singular at $r=0$.  There are techniques to ``regularise'' the collision, leading to a continuation of the trajectory after collision, but for this paper we will ignore the issue.

There is an alternative Hamiltonian formulation with modified momentum $\bm{\pi} = \bm{p}+\bm{A}$, $\omega = \sum_i dr_i \wedge d\pi_i$, $H = \gamma - \frac{1}{r}$, where $\gamma = \sqrt{1+|\bm{\pi}-\bm{A}|^2}$, but we prefer to put the magnetic effect into the symplectic form and keep the standard momentum $\bm{p}$.  

There are also formulations in space-time with respect to proper time.  The simplest one has the feature that the ``electrostatic'' part of gravity is also put into the symplectic form. 
We denote by $Q=(t,\bm{r})$ the position of the particle in Minkowski space, with inner product $Q \cdot Q' = -tt' + \bm{r} \cdot \bm{r'}$ and associated ``norm''-squared $|Q|^2 = Q \cdot Q$.
It has Hamiltonian $K = \frac12|P|^2$, where $P = (-\gamma,\bm{p})$ is the 4-momentum, and symplectic form $\Omega = -d\Theta - F$, where $\Theta = \sum_\nu P_\nu dQ^\nu$ is the canonical 1-form on the tangent bundle of Minkowski space and $F = F_{\mu\nu}dQ^\mu \wedge dQ^\nu$ is the Faraday tensor, which has components
$$F_{\mu\nu} = \left[\begin{array}{cccc} 
0 & E_x & E_y & E_z \\
-E_x & 0 & -B_z & B_y \\
-E_y & B_z & 0 & - B_x \\
-E_z & -B_y & B_x & 0 
\end{array}\right].$$
This Hamiltonian system must be restricted to the level set $K=-\frac12$ (for unit rest mass).
An alternative is to put the gravitational fields into the Hamiltonian and use the canonical symplectic form; we will mention a version of that in the next section.

In Hamiltonian dynamics, there is a natural relation between continuous symmetries and constants of the motion.  In particular,
we can check that the conserved quantity $L_z$ follows from $i_{\partial_\phi}H=0$ and $i_{\partial_\phi}\omega = dL_z$.  We use
\begin{equation}
\omega = -d\Theta - dA^\flat,
\end{equation}
with the natural 1-form $\Theta = p_r dR + p_z dz + Rp_\phi d\phi$,
and note that 
\begin{equation}
dA^\flat = \tfrac{D}{r^5} (2r^2-3R^2) R\, dR \wedge d\phi + 3\tfrac{D}{r^5} R^2z\, dz \wedge d\phi.
\end{equation}
So 
\begin{equation}
\omega = dR \wedge dp_R + dz \wedge dp_z + d\phi \wedge d(Rp_\phi )- dA^\flat,
\end{equation}
and thus 
\begin{equation}
i_{\partial_\phi}\omega = d(Rp_\phi)+\tfrac{D}{r^5} (2r^2-3R^2) R\, dR + 3 \tfrac{D}{r^5} R^2 z\, dz.
\label{eq:21}
\end{equation}
For comparison, from (\ref{eq:Lz}),
\begin{equation}
dL_z = d(Rp_\phi) - D\, d\left(\tfrac{R^2}{r^3}\right),
\end{equation}
the second term of which can be expanded as
$\frac{D}{r^5} \left( 2R r^2 dR - 3R^3 dR - 3R^2z\, dz\right)$, giving a result that agrees with (\ref{eq:21}).

\section{Reduced system}

The invariance of the Hamiltonian structure under rotation about the $z$-axis leads to a reduced Hamiltonian system of 2 DoF, in just $(R,z,p_R,p_z)$.

The reduced equations of motion for given value of $L_z$ can be obtained by expressing $H$ and $\omega$ in cylindrical coordinates
and substituting (from (\ref{eq:Rpphi}))
\begin{equation}
p_\phi = \frac{L_z}{R} - \frac{DR}{(R^2+z^2)^{3/2}},
\label{eq:pphi}
\end{equation}
leading to
\begin{equation}
H = \sqrt{1+p_R^2+p_z^2+\left(\tfrac{L_z}{R}-\tfrac{DR}{(R^2+z^2)^{3/2}}\right)^2} - \tfrac{1}{\sqrt{R^2+z^2}},
\label{eq:Ham}
\end{equation}
\begin{equation}
\omega = dR \wedge dp_R + dz \wedge dp_z.
\label{eq:symp}
\end{equation}
As the symplectic form is canonical, the equations of motion are in canonical form: 
\begin{eqnarray}
\dot{R} &=& \tfrac{p_R}{\gamma} \\
\dot{z} &=& \tfrac{p_z}{\gamma} \label{eq:dotz} \\
\dot{p}_R &=& \tfrac{p_\phi}{\gamma} \left( \tfrac{L_z}{R^2}+\tfrac{D}{r^3}-\tfrac{3DR^2}{r^5}\right) - \tfrac{R}{r^3} \\
\dot{p}_z &=& -\tfrac{p_\phi}{\gamma} \tfrac{3DRz}{r^5} - \tfrac{z}{r^3} , \label{eq:dotpz}
\end{eqnarray}
with $p_\phi$ given by (\ref{eq:pphi}) and $\gamma$ the first square root in (\ref{eq:Ham}).
If $L_z \ne 0$, the reduction appears to have introduced a singularity on the whole of the $z$-axis $R=0$, but conservation of $H$ implies that the only way that $R=0$ can be reached is if $r \to 0$.
The motion in $\phi$ can be reconstructed by $\dot{\phi} = \frac{p_\phi}{\gamma R}$.

The system can be written in an alternative form with respect to proper time $\tau$ for the test particle.  Namely, the motion for $H=E$ is the motion in proper time for Hamiltonian
\begin{equation}
J = \tfrac12 \left(p_R^2+p_z^2 - (E+\tfrac{1}{r})^2 + 1 + (\tfrac{L_z}{R}-\tfrac{DR}{r^3})^2\right)
\label{eq:J}
\end{equation}
on $J=0$, with respect to the canonical symplectic form (\ref{eq:symp}), as can be checked by explicit comparison, but one must restrict to $E+\frac{1}{r}\ge 1$.

To analyse the dynamics of the reduced system it is convenient first to notice that the equatorial plane $z=0, p_z=0$ is invariant.  So we start by analysing the dynamics there.

\section{Equatorial motion}
\label{sec:equator}

The plane $z=0, p_z=0$ is invariant for any value of $D$.  
The motion in it is a 1DoF Hamiltonian system in just $(R,p_R)$.  It has
$$H = \sqrt{1+p_R^2+\left(\tfrac{L_z}{R}-\tfrac{D}{R^2}\right)^2} - \tfrac{1}{R}, \quad \omega = dR \wedge dp_R.$$
We see that for given value $E$ of $H$ and values of $L_z, D$, 
the allowed region of $R$ is 
\begin{equation}
\left(E+\tfrac{1}{R}\right)^2 \ge 1 + \left(\tfrac{L_z}{R} - \tfrac{D}{R^2}\right)^2,
\end{equation} 
with $R> 0$ and $\frac{1}{R}\ge 1-E$.  As the expression is a polynomial of degree at most 4 in $1/R$, this consists of one or two intervals or is empty.
  
\begin{figure}[htbp]
    \centering
    \begin{subfigure}[b]{0.3\textwidth}
        \centering
        \includegraphics[height=2.0in]{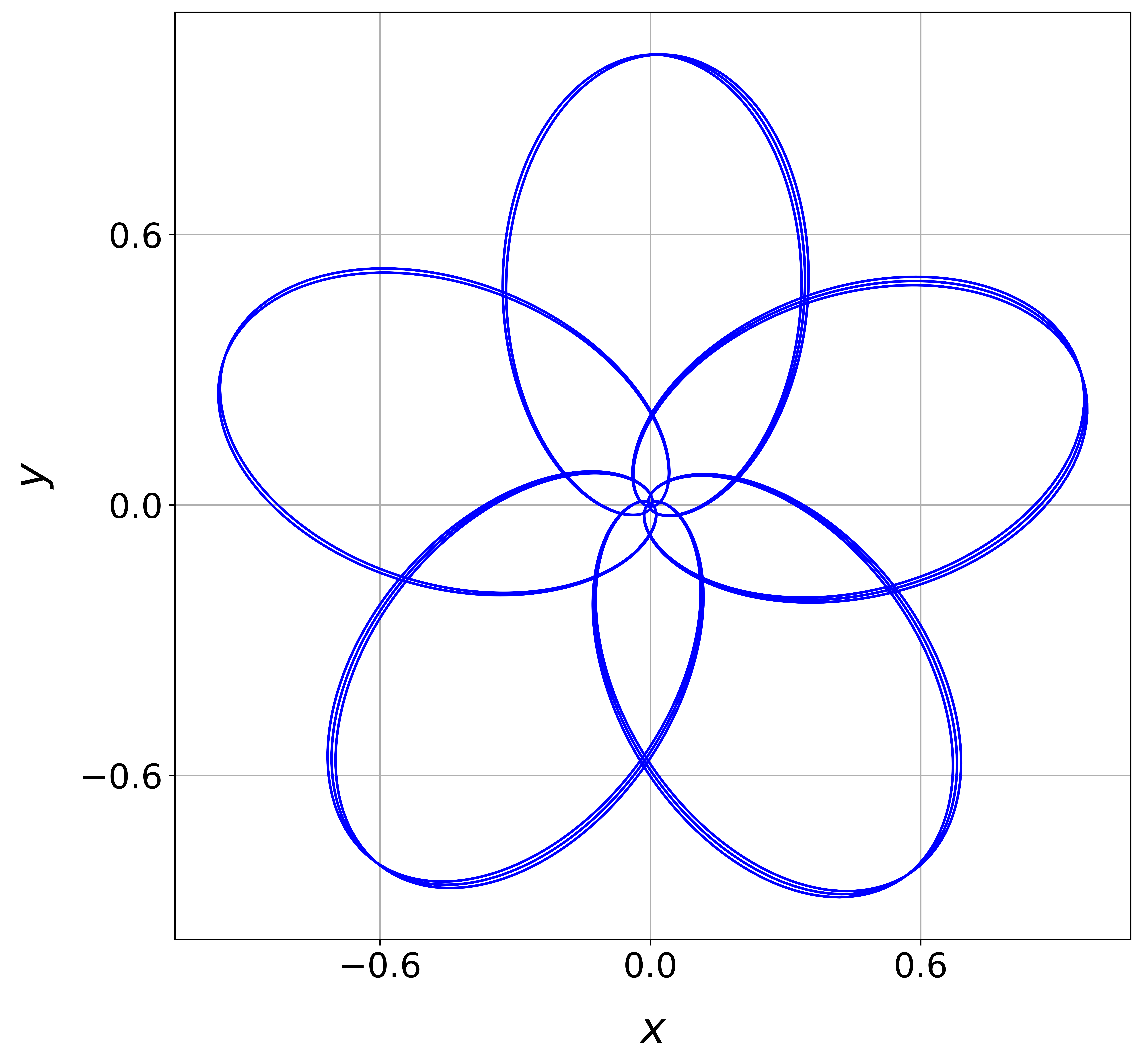}
        \caption{$(E, L_z, D)=(0.118, -0.500, 1.00\times10^{-4})$}
    \end{subfigure}
    \hfill
    \begin{subfigure}[b]{0.3\textwidth}
        \centering
        \includegraphics[height=2.0in]{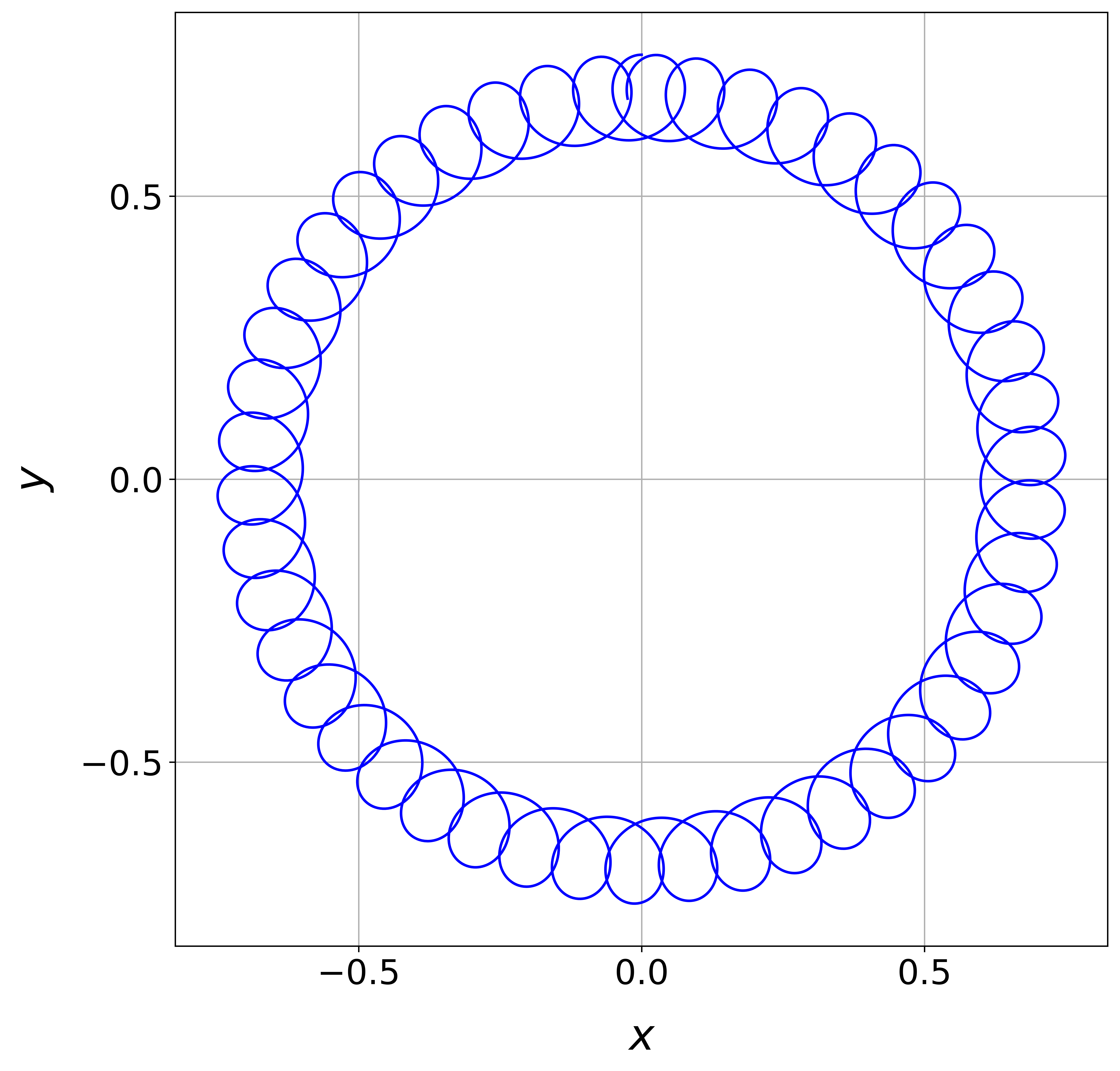}
        \caption{$(E, L_z,D)=(-0.215, 4.38, 3.00)$}
    \end{subfigure}
    \hfill
    \begin{subfigure}[b]{0.3\textwidth}
        \centering
        \includegraphics[height=2.0in]{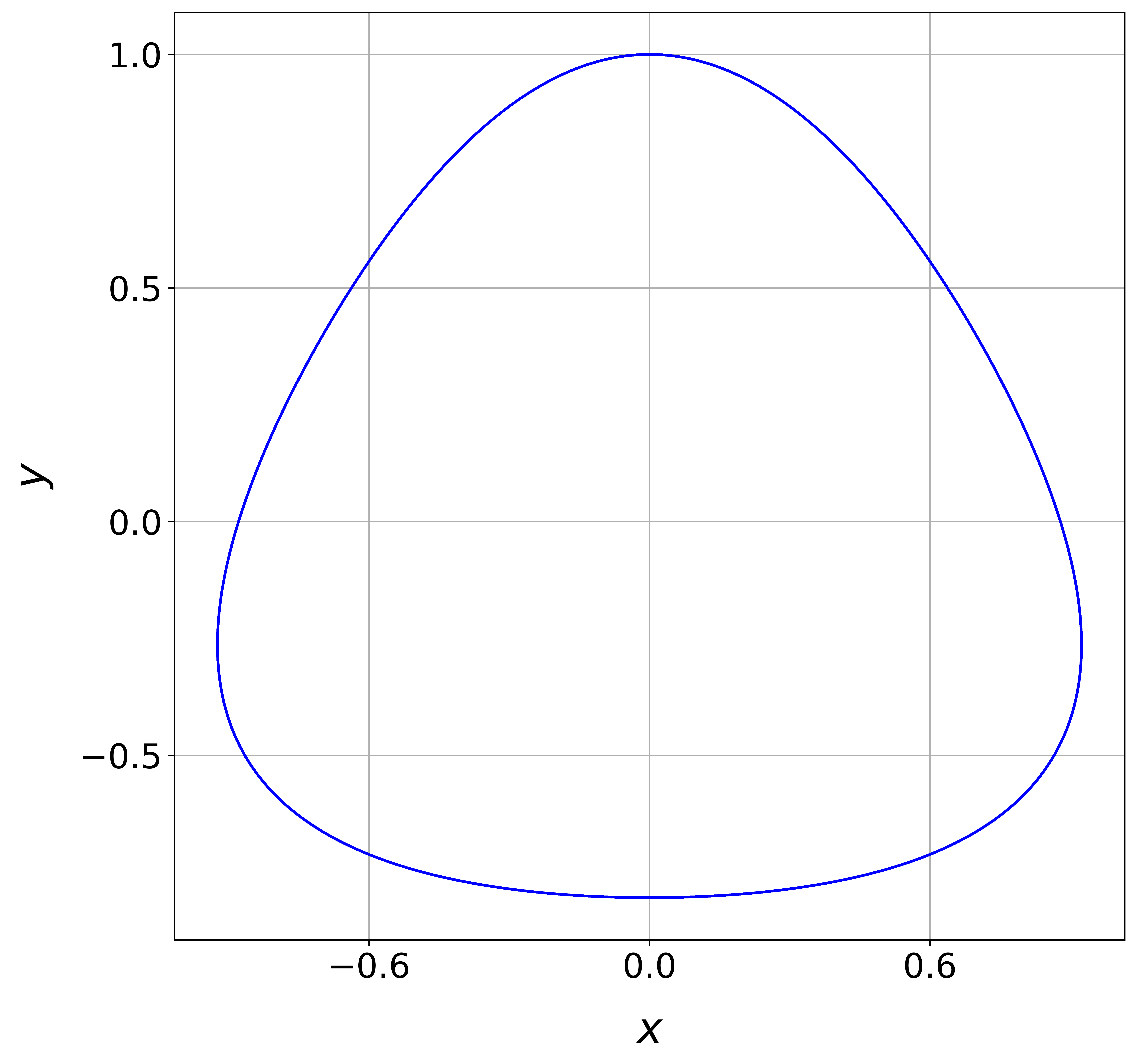}
        \caption{ $(E,L_z,D)=(0.118, 0.500, 1.00)$}
    \end{subfigure}
    \caption{Equatorial orbits showing a) precession, b) gyration and c) periodicity.}
    \label{fig:equatorialorbits}
\end{figure}

In each interval, the motion is periodic between the endpoints, except if an endpoint is at $R=0$ or $R=\infty$.  In the latter case one is obliged to take $E\ge 1$ and it takes infinite time to reach it.  The former case is impossible if $D>0$ because $D/R^2$ beats $1/R$ as $R \to 0$.  
The motion of the original system resulting from periodic motion of $R$ in an interval $[a,b]$ with $a>0$ resembles a precessing ellipse, that may surround the origin or not, as can be seen in Figure~\ref{fig:equatorialorbits}.
The precession rate goes to $0$ in the non-relativistic limit for $D=0$ (Kepler ellipses). 

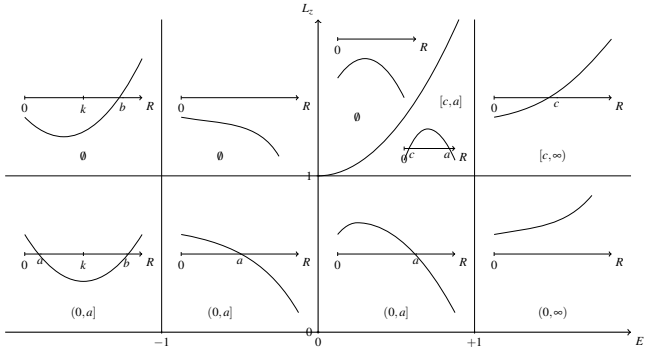
\begin{figure}[htbp] 
    \centering
\scalebox{0.52}{
\begin{tikzpicture}
\draw[thick,->] (-8,0) -- (8,0) node[anchor=north west]{$E$};
\foreach \x in {-1,0,+1} \draw (4*\x cm,1pt) -- (4*\x cm, -1pt) node[anchor=north]{$\x$};
\draw (-8,4) -- (8,4);
\draw[thick,->] (0,0) -- (0,8) node[anchor=south east]{$L_z$};
\foreach \y in {0,1} \draw (1pt,4*\y cm) -- (-1pt,4*\y cm) node[anchor=east]{$\y$};
\draw (-4,0) -- (-4,8);
\draw (4,0) -- (4,8);
\draw (0,4) parabola (3.6,8);

\draw[->] (-7.5,2) -- (-4.5,2) node[anchor=north west] {$R$};
\draw (-7.5,2.5) parabola bend (-6,1.3) (-4.5,2.5);
\draw (-7.5,2.05) -- (-7.5,1.95) node[anchor=north]{$0$};
\draw (-7.1,2) -- (-7.1,2) node[anchor=north]{$a$};
\draw (-6,2.05) -- (-6,1.95) node[anchor=north]{$k$};
\draw (-4.9,2) -- (-4.9,2) node[anchor=north]{$b$};
\node[] at (-6,0.5) {$(0,a]$};

\draw[->] (-7.5,6) -- (-4.5,6) node[anchor=north west] {$R$};
\draw (-7.5,5.5) parabola bend (-6.5,5) (-4.5,7);
\draw (-7.5,6.05) -- (-7.5,5.95) node[anchor=north]{$0$};
\draw (-6,6.05) -- (-6,5.95) node[anchor=north]{$k$};
\draw (-5,6) -- (-5,6) node[anchor=north]{$b$};
\node[] at (-6,4.5) {$\emptyset$};

\draw[->] (-3.5,2) -- (-0.5,2) node[anchor=north west]{$R$};
\draw (-3.5,2.5) to[out=-10,in=120] (-0.5,0.5);
\draw (-3.5,2.05) -- (-3.5,1.95) node[anchor=north]{$0$};
\draw (-2,2) -- (-2,2) node[anchor=north]{$a$};
\node[] at (-2.5,0.5) {$(0,a]$};

\draw[->] (-3.5,6) -- (-0.5,6) node[anchor=north west]{$R$};
\draw (-3.5,5.5) to[out=-10,in=120] (-1,4.5);
\draw (-3.5,6.05) -- (-3.5,5.95) node[anchor=north]{$0$};
\node[] at (-2.5,4.5) {$\emptyset$};

\draw[->] (0.5,2) -- (3.5,2) node[anchor=north west]{$R$};
\draw (0.5,2.5) parabola bend (1,2.8) (3.5,0.5);
\draw (0.5,2.05) -- (0.5,1.95) node[anchor=north]{$0$};
\draw (2.5,2) -- (2.5,2) node[anchor=north]{$a$};
\node[] at (2,0.5) {$(0,a]$};

\draw[->] (4.5,2) -- (7.5,2) node[anchor=north west]{$R$};
\draw (4.5,2.5) to[out=10,in=-130] (7,3.5);
\draw (4.5,2.05) -- (4.5,1.95) node[anchor=north]{$0$};
\node[] at (6,0.5) {$(0,\infty)$};

\draw[->] (4.5,6) -- (7.5,6) node[anchor=north west]{$R$};
\draw (4.5,5.5) to[out=10,in=-130] (7.5,7.5);
\draw (4.5,6.05) -- (4.5,5.95) node[anchor=north]{$0$};
\draw (6.1,6) -- (6.1,6) node[anchor=north]{$c$};
\node[] at (6,4.5) {$[c,\infty)$};

\draw[->] (0.5,7.5) -- (2.5,7.5) node[anchor=north west]{$R$};
\draw (0.5,6.5) parabola bend (1.2,7) (2.2,6);
\draw (0.5,7.55) -- (0.5,7.45) node[anchor=north]{$0$};
\node[] at (1,5.5) {$\emptyset$};

\draw[->] (2.2,4.7) -- (3.5,4.7) node[anchor=north west]{$R$};
\draw (2.2,4.4) parabola bend (2.8,5.2) (3.5,4.4);
\draw (2.2,4.75) -- (2.2,4.65) node[anchor=north]{$0$};
\draw (2.4,4.7) -- (2.4,4.7) node[anchor=north]{$c$};
\draw (3.3,4.7) -- (3.3,4.7) node[anchor=north]{$a$};
\node[] at (3.4,5.9) {$[c,a]$};

\end{tikzpicture}
}
\caption{Allowed intervals in $R$ for equatorial motion with $D=0$ in various regions of the $(E,L_z)$-plane. $\emptyset$ signifies that there is no allowed interval.  The results are independent of the sign of $L_z$ so we plot for only $L_z\ge 0$. The curve in the parameter space is $(1-E^2)L_z^2 = 1$.  The insets show sample graphs of $(E^2-1)R^2+2ER + (1-L_z^2)$ as a function of $R>0$.  Note that the constraint $R \le k = 1/(1-E)$ removes the apparent unbounded interval for $E<-1$.
Note also that $a \to \infty$ as $E \nearrow 1$, $b \to \infty$ as $E \nearrow -1$, $a \to 0$ as $L_z \nearrow 1$ for $E<0$, and $c\to 0$ as $L_z \searrow 1$ for $E>0$.
}
    \label{fig:D=0}
 \end{figure}

To find the allowed intervals, we treat first the case $D=0$.  Then the allowed region is given by
\begin{equation}
(E^2-1)R^2 + 2{E}{R} + (1-L_z^2) \ge 0,
\label{eq:allowed}
\end{equation}
with $R\ge 0$ and 
\begin{equation}
R \le k = \tfrac{1}{1-E}.
\label{eq:defk}
\end{equation}
A priori, this is one or two intervals or empty.  The separating cases are when there is a double root $(1-E^2)L_z^2 = 1$ or an endpoint moves through $R=0$ or $\infty$ ($L_z^2=1$ or $E^2=1$, respectively).  The case $R=\infty$ with $E=-1$ is ruled out by the constraint $R \le k$, so the case of an endpoint moving through $R=\infty$ is possible for only $E=+1$.
The result is shown in Figure~\ref{fig:D=0}.
We see that in fact there is at most one interval, because at $R= k$ the function in (\ref{eq:allowed}) has the value $-L_z^2 \le 0$.

Note that this analysis of equatorial motion in the case $D=0$ actually applies to the whole dynamics for $D=0$.
This is because the third constant of the motion restricts the motion to the plane perpendicular to the initial ${\bf L} = {\bf r} \times {\bf p}$.  By a rotation of axes, we can take it to be the equatorial plane.  

Turning to $D>0$, we already remarked that $R=0$ is excluded. 
So the separating cases for the number of allowed intervals are associated with critical points of $H$ or passage of an endpoint through $R=\infty$.  Just as for $D=0$, passage of an endpoint through $\infty$ requires $E=+1$.
The critical points correspond precisely to the equatorial circular orbits, because the critical points are the equilibria of the reduced dynamics and we have restricted to $z=0$.

To study the equatorial circular orbits in more detail, it is tidier to write 
\begin{equation}
u=\tfrac{1}{R}.
\end{equation}
Then the equations for critical points are
\begin{eqnarray}
(E+u)^2 &=& 1 + (L_zu-Du^2)^2 \label{eq:endpt} \\
E+u &=& (L_zu-Du^2)(L_z-2Du) .
\end{eqnarray}
We can write $E+u = \gamma$ and $L_zu-Du^2 = p_\phi$.  So the second equation is 
\begin{equation}
\gamma u = p_\phi(p_\phi-Du^2),
\end{equation}
 which is the same as (\ref{eq:circeq}).  It has solutions
\begin{equation}
u = \frac{-\gamma +\sigma \sqrt{\gamma^2+4D p_\phi^3}}{2Dp_\phi}
\label{eq:u}
\end{equation}
with $\sigma \in \{\pm\}$.

Then the number of allowed intervals changes by $\pm 1$ on crossing a curve of equatorial circular orbit in Figure~\ref{fig:circ}.
By checking suitable cases, the number of intervals is two in the region bounded by the cusped curve, one in the region between the two curves, and zero to the left of the smooth curve.

As promised in Section~\ref{sec:eqcirc}, we now explain the shapes of the curves in Figure~\ref{fig:circ}.  They are smooth curves, except at a triple root of (\ref{eq:endpt}), where the curve can be expected to form a semi-cubic cusp in the parameter space. The condition for a triple root is
\begin{equation}
1 = (L_z-2Du)^2-2D(L_zu-Du^2),
\end{equation}
which can be written as 
\begin{equation}
1 = u^{-2}(p_\phi-Du^2)^2-2Dp_\phi .
\end{equation}
Combining with the equation for a double root, we obtain
$2D p_\phi^3 = 1$, i.e.~
\begin{equation}
p_\phi = p_* = (2D)^{-1/3},
\label{eq:defpstar}
\end{equation}
and  $\gamma = \sqrt{1+p_*^2}$.

Using (\ref{eq:u}), we obtain that the triple root is at
\begin{equation}
u = p_*^2 \left(\sqrt{3+p_*^2}-\sqrt{1+p_*^2}\right),
\end{equation}
taking $\sigma=+$ because we require $u>0$.  Then we obtain that the position of the cusp is at 
\begin{equation}
E = \gamma-u = (1+p_*^2)^{3/2} - p_*^2\sqrt{3+p_*^2}
\end{equation}
and
\begin{equation}
L_z = Du + \tfrac{p_\phi}{u} = 
\sqrt{1+3p_*^{-2}}.
\end{equation}
Note that this has $L_z>1$.  It also has $E<1$, because writing $x = p_*^2$ we have $E = (1+x)^{3/2}-x(3+x)^{1/2}$, so at $x=0$ we have $E=1$, and $$\tfrac{dE}{dx} = \tfrac32 (1+x)^{1/2} - (3+x)^{1/2} -\tfrac12 x (3+x)^{-1/2} < 0,$$ because $(2+x)^2 = 1 + (1+x)(3+x)$.
In principle, one could check that the generic conditions for a semi-cubic cusp are satisfied, but we did not do it.

\section{Analysis of 2DoF reduced dynamics}

We see from (\ref{eq:Ham}) or (\ref{eq:J}) that for energy $H=E$,
\begin{equation}
p_z^2+p_R^2 = K := \left(E+\tfrac{1}{r}\right)^2 - 1 -\left(\tfrac{L_z}{R}-\tfrac{DR}{r^3}\right)^2.
\label{eq:Hill}
\end{equation}
This has to be at least 0, so the motion is restricted to the region of $(R,z)$ where $K\ge 0$.  Note that the motion is also restricted to $\frac{1}{r} \ge 1-E$ because the first square root in (\ref{eq:Ham}) is positive and its argument is at least 1.
We call the result of these two restrictions {\em Hill's region}, by analogy with celestial mechanics.  

\begin{figure*}[t] 
  \centering
  
  \begin{subfigure}[t]{0.66\textwidth} 
    \centering
    \includegraphics[height=1.8in]{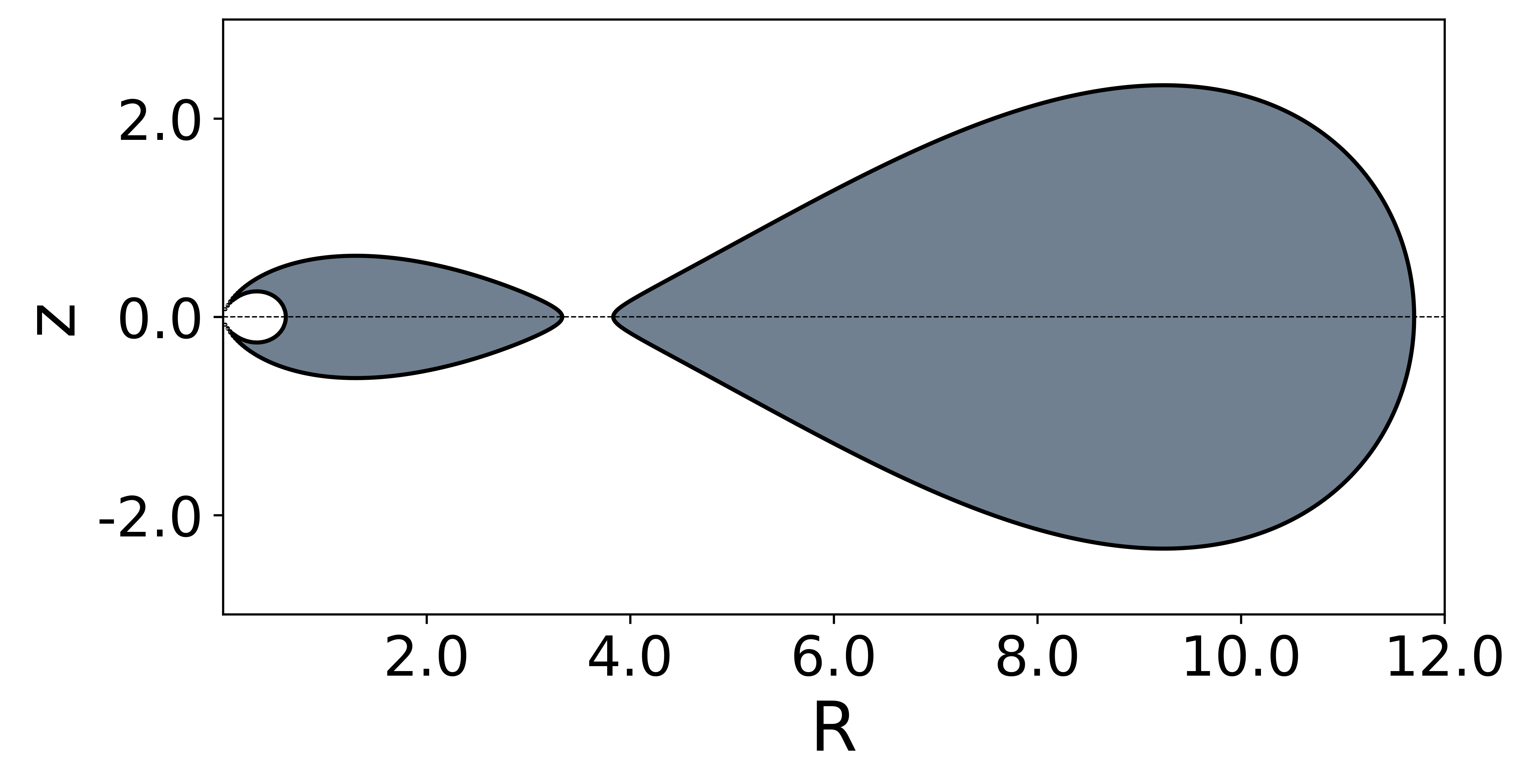}
    \caption{$(0.950, 3.40, 3.00)$}
  \end{subfigure}
  \hfill
  \begin{subfigure}[t]{0.33\textwidth}
    \centering
    \includegraphics[height=1.8in]{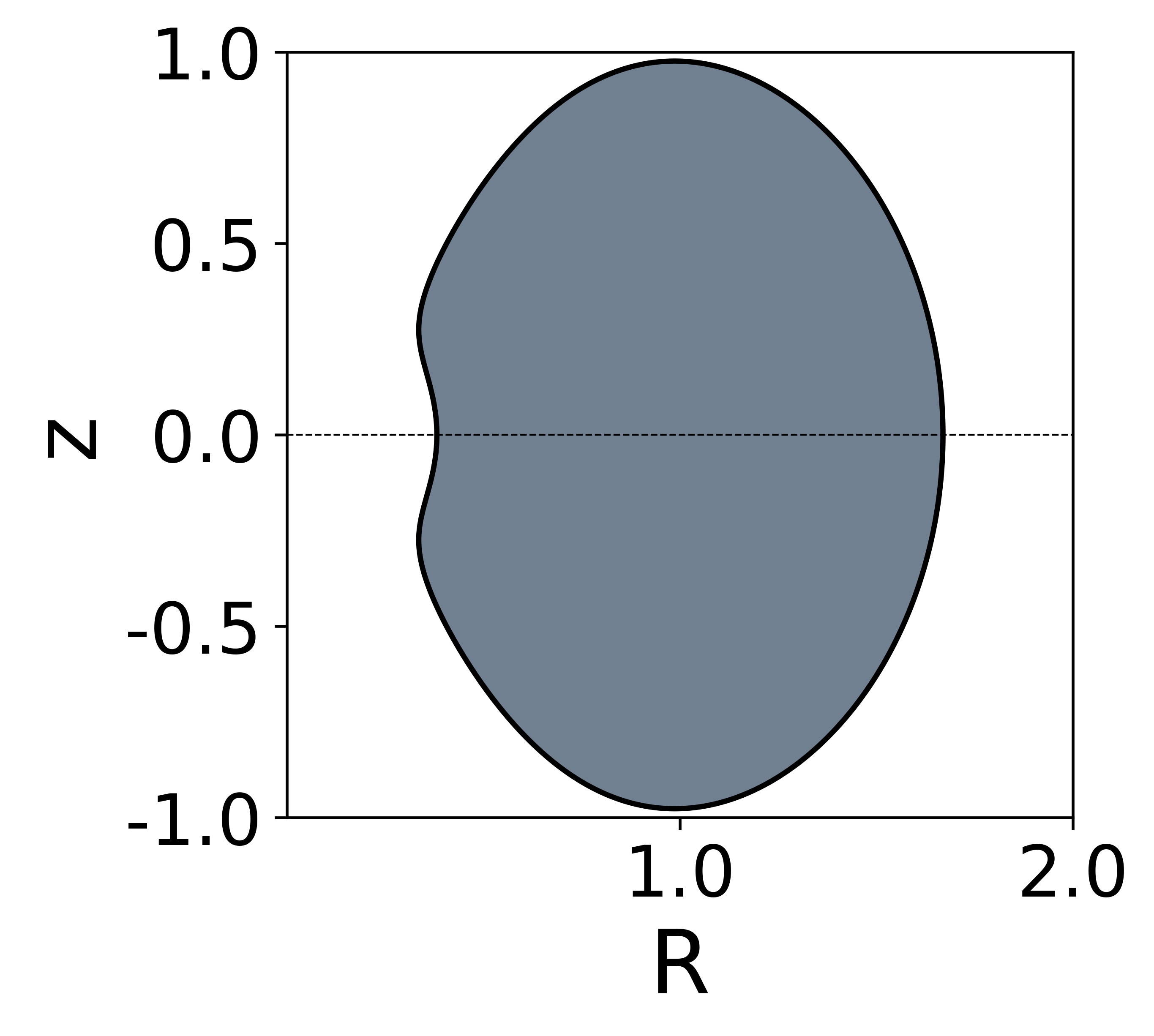}
    \caption{$(0.480, -0.600, 0.200)$}
  \end{subfigure}
  
  \vspace{1em} 
  
  \begin{subfigure}[t]{0.33\textwidth}
    \centering
    \includegraphics[height=1.8in]{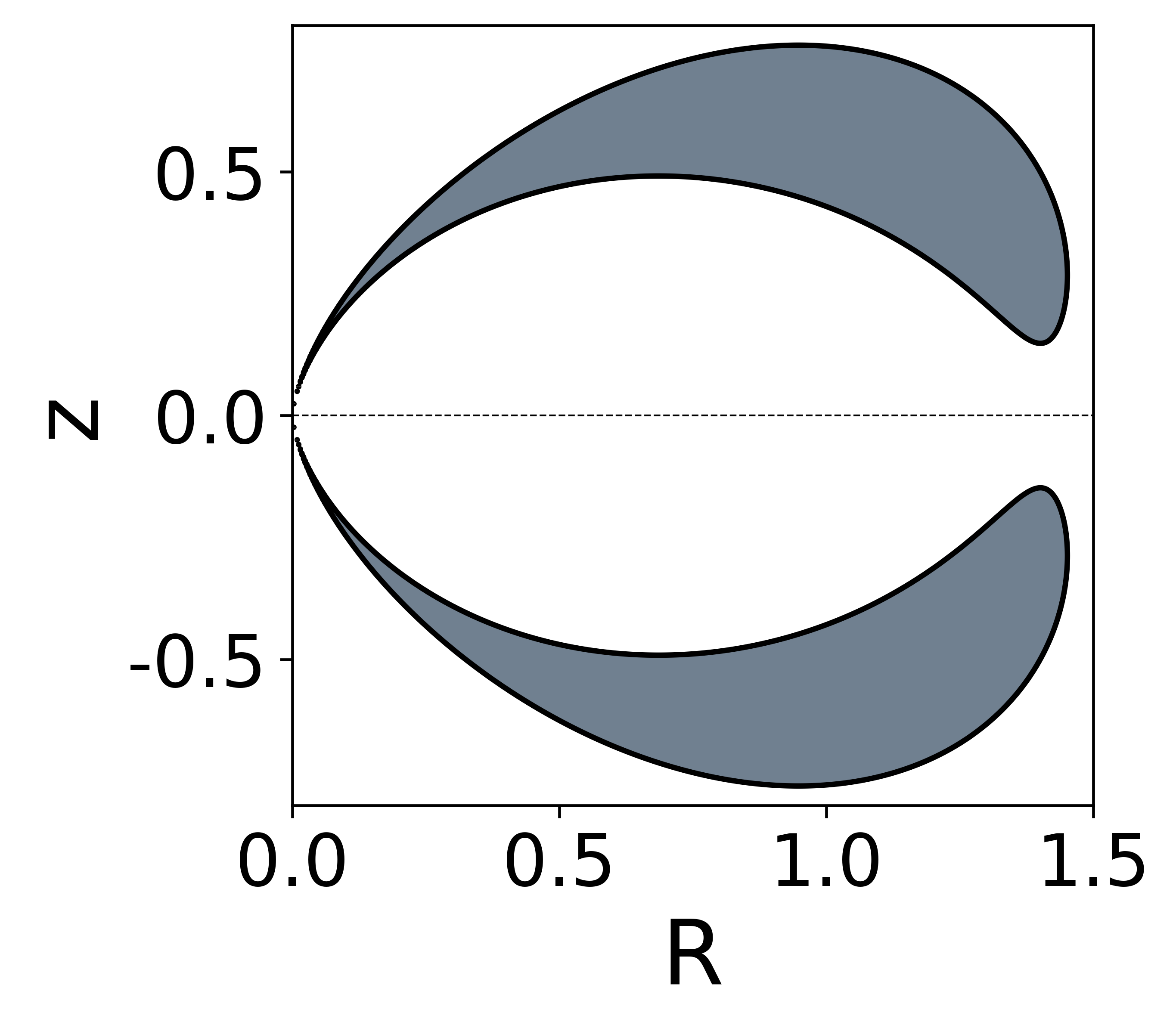}
    \caption{$(0.330, 2.90, 4.70)$}
  \end{subfigure}
  \hfill
  \begin{subfigure}[t]{0.33\textwidth}
    \centering
    \includegraphics[height=1.8in]{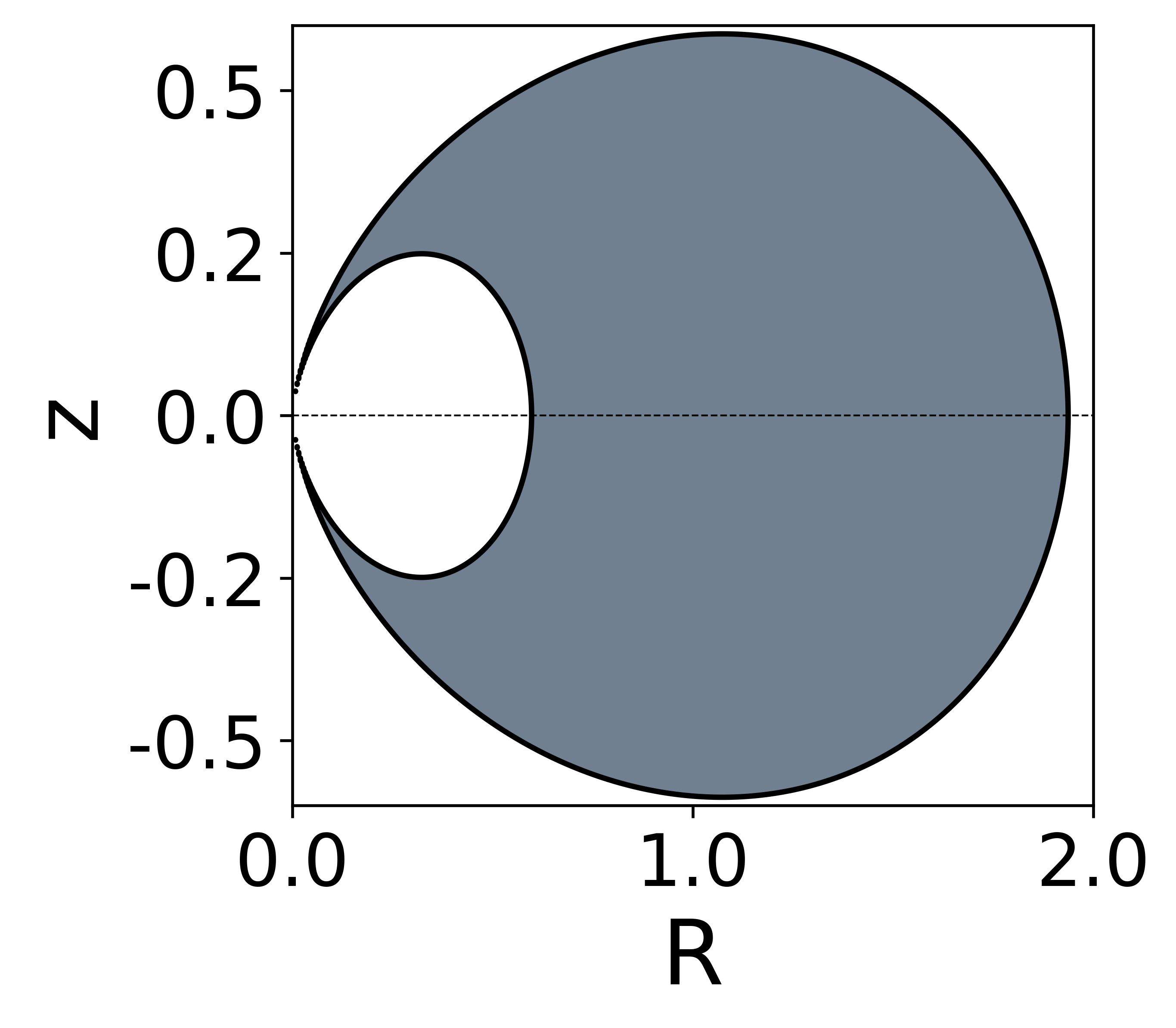 }
    \caption{$(0.800, 3.00, 2.600)$}
  \end{subfigure}
  \hfill
  \begin{subfigure}[t]{0.33\textwidth}
    \centering
    \includegraphics[height=1.8in]{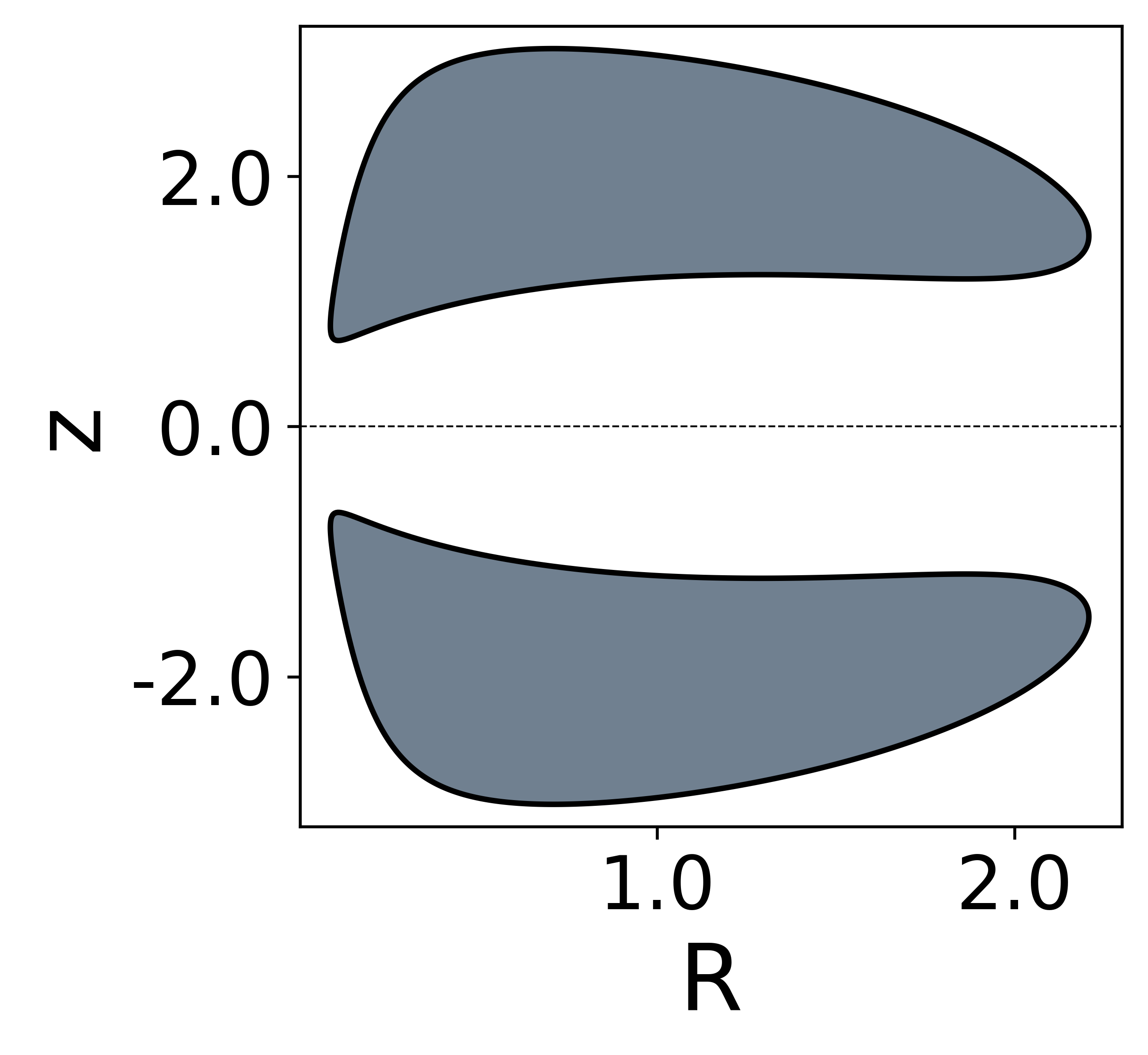}
    \caption{$(0.700, -0.100, 3.00)$}
  \end{subfigure}

  \vspace{1em}
  \begin{subfigure}[t]{0.33\textwidth}
    \centering
    \includegraphics[height=1.8in]{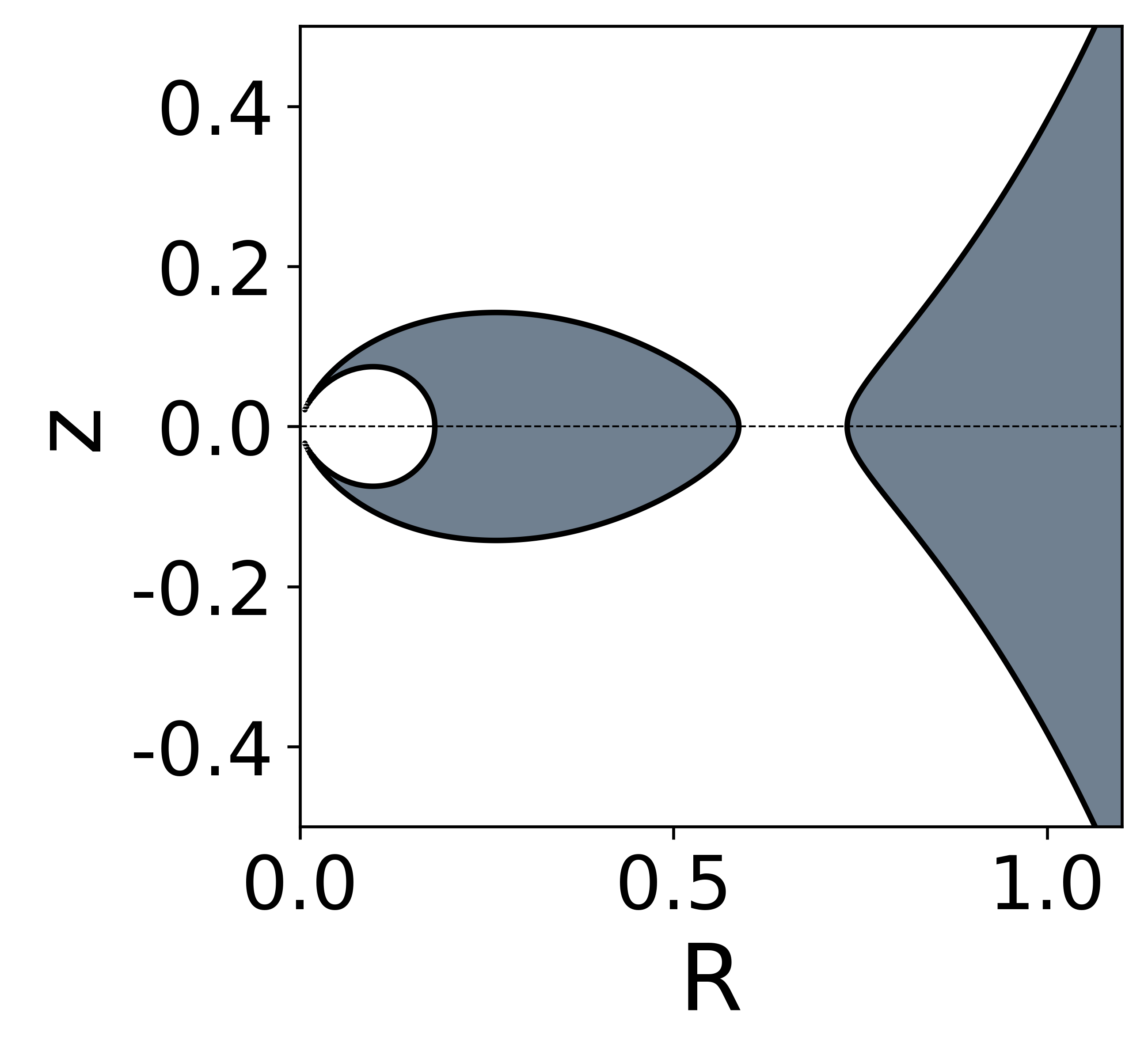}
    \caption{$(2.500, 4.10, 1.00)$}
  \end{subfigure}
  \hfill
  \begin{subfigure}[t]{0.33\textwidth}
    \centering
    \includegraphics[height=1.8in]{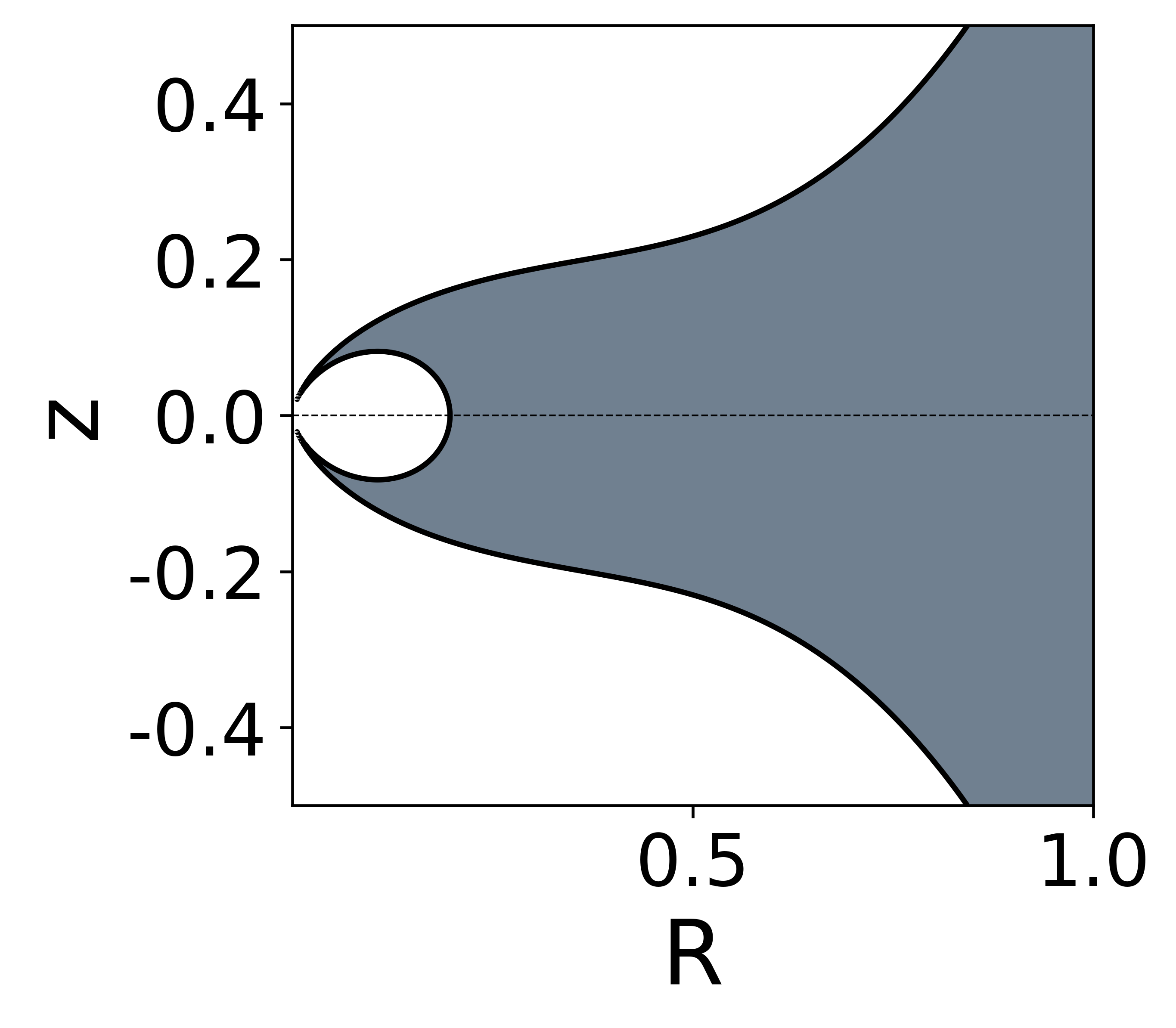}
    \caption{$(2.50, 3.60,1.00)$}
  \end{subfigure}
  \hfill
  \begin{subfigure}[t]{0.33\textwidth}
    \centering
    \includegraphics[height=1.8in]{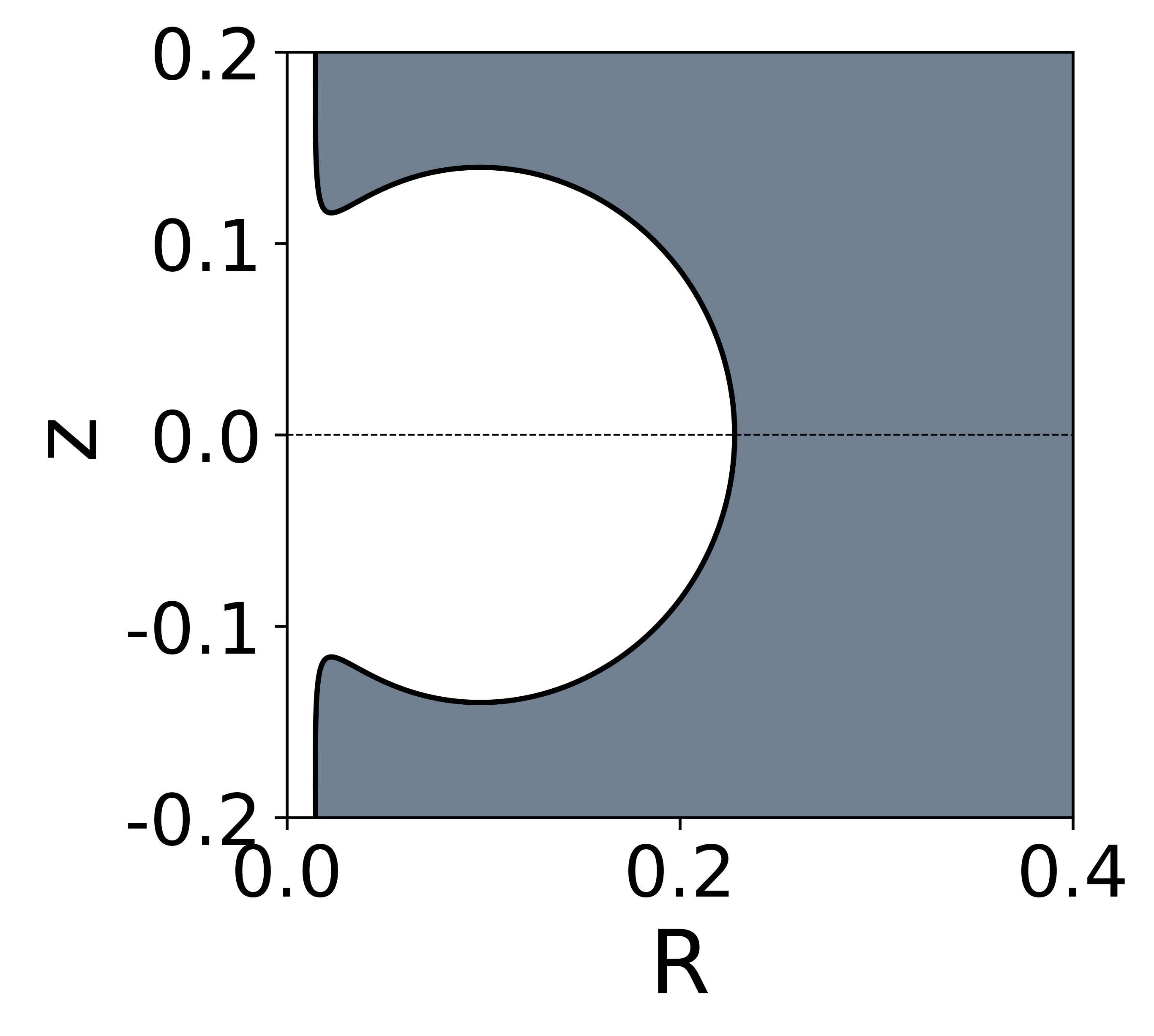}
    \caption{$(10.0, -0.200, 0.700)$}
  \end{subfigure}
  
  \caption{Hill's regions for various choices of $(E, L_z, D)$. Negative $L_z$ components are on the right most column.}
  \label{fig:Hill}
\end{figure*}

Figure~\ref{fig:Hill} shows examples of the Hill's region for a variety of choices of the parameters $(E,L_z,D)$.  We take $D>0$ in the following discussion. 

For $L_z\le 0$, the $(DR/r^3)^2$ term in (\ref{eq:Hill}) beats the $1/r^2$ term as $r \to 0$, so non-negativity of $p_R^2+p_z^2$ implies $r=0$ is not accessible.
In contrast, for $L_z>0$, the Hill's region has horns around a ``centre curve'' $L_z/R = DR/r^3$ that extends all the way to $r=0$.  The equation for the centre curve can be written as $L_z^2 (R^2+z^2)^3 = D^2 R^4$.  These observations agree with the figures.

We see from the figures that the Hill's region may have one or two components.  Also, but not shown, for some choices of parameters it is empty.  
As the parameters are varied, the number of components can change only when the boundary contains a critical point of $K$ (for fixed $E,L_z,D$) with $K=0$, or when a component crosses $R=0$ or $R=\infty$.  We concentrate on the case of a critical point. 

For a local maximum or minimum of $K$ with $K=0$, the component is a single point and a transition can occur as parameters are varied, in which a component is created or destroyed.  For a saddle of $K$ with $K=0$, two components (locally) can merge or one component separate into two (locally) as parameters are varied.

We compute the equations for the critical points of $K$ by differentiating $K$ with respect to $z$ and $R$, respectively:
\begin{eqnarray}
z\left((E+\tfrac{1}{r})r^{-3} + (\tfrac{L_z}{R}-\tfrac{DR}{r^3})\tfrac{3DR}{r^5}\right) &=& 0 \\
\left(E+\tfrac{1}{r}\right)\tfrac{R}{r^3} + \left(\tfrac{L_z}{R}-\tfrac{DR}{r^3}\right)\left(\tfrac{3DR^2}{r^5} - \tfrac{L_z}{R^2}-\tfrac{D}{r^3}\right) &=& 0
\end{eqnarray}
The first equation gives $z=0$ or its second factor zero.  
We already treated the equatorial case $z=0$ in the previous section.


For the non-equatorial critical points, writing $p_\phi = \tfrac{L_z}{R}-\tfrac{DR}{r^3}$ and $\gamma = E+\tfrac{1}{r}$, the equations are
\begin{eqnarray}
\gamma^2 &=& 1 + p_\phi^2 \\
\gamma r^2 + 3p_\phi DR &=& 0 \\
\gamma \tfrac{R}{r^3} + p_\phi (\tfrac{3DR^2}{r^5}-\tfrac{L_z}{R^2}-\tfrac{D}{r^3}) &=& 0 .
\end{eqnarray}
Writing $v_\phi = p_\phi/\gamma$, it follows from the second that 
\begin{equation}
v_\phi = -\frac{r^2}{3DR} < 0, 
\end{equation}
and inserting $p_\phi = \gamma v_\phi$ into the third, we obtain
\begin{equation}
L_z r^3 = -DR^2, 
\end{equation}
so in particular $L_z<0$.
Then 
\begin{equation}
p_\phi = \frac{L_z}{R}-\frac{DR}{r^3} = - 2\frac{DR}{r^3}.
\end{equation}
Recalling (\ref{eq:gamma}), $\gamma$ can be written in two ways:
\begin{equation}
\sqrt{1+ 4\frac{D^2R^2}{r^6}} =\gamma= 1/\sqrt{1-9\frac{r^4}{D^2R^2}}.
\end{equation}
It follows that
\begin{equation}
4\frac{D^2R^2}{r^6} - \frac{r^4}{9D^2R^2} - \frac{4}{9r^2} = 0.
\end{equation}
So we can parametrise the set of non-equatorial critical points by $v_\phi=v$:
\begin{eqnarray}
r^2 &=& \tfrac49 (v^{-4}-v^{-2}) \\
R &=& -\tfrac{4}{27 D} (v^{-5}-v^{-3}).
\end{eqnarray}
It is necessary, however, to require $z^2=r^2-R^2\ge 0$.  This restricts $v$ to the set where
\begin{equation}
\tfrac49 (v^{-4}-v^{-2}) \ge \tfrac{16}{3^6 D^2} (v^{-5}-v^{-3})^2,
\end{equation}
i.e.~$v^6 \ge \frac{4}{81D^2}(1-v^2)$, equivalently (remembering that $v<0$), $\gamma v^3 <-\frac{2}{9D}$. 
To complete the parametric representation, 
\begin{eqnarray}
E &=& \gamma - \frac{1}{r} = \frac{1-\tfrac32 v^2}{\sqrt{1-v^2}}, \\
L_z &=& -\frac{DR^2}{r^3} = -\frac{2}{27Dv^4} \sqrt{1-v^2}.
\end{eqnarray}

\begin{figure}[htbp] 
   \centering
\includegraphics[width=3in]{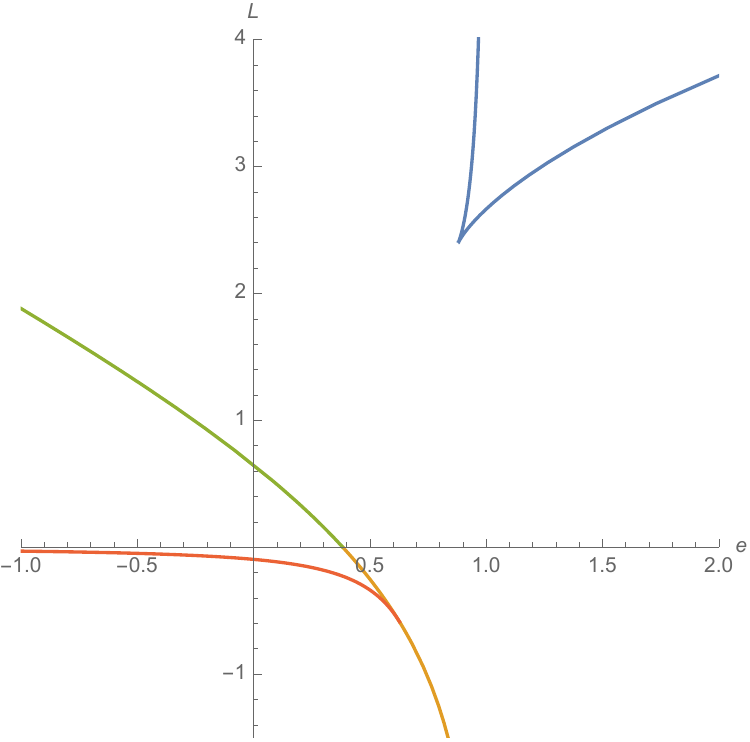} 
   \caption{The curves of all critical points of $H$ as functions of $(E,L_z)$ for $D=1$. }
   \label{fig:critpts}
\end{figure}

In Figure~\ref{fig:critpts} this curve of non-equatorial circular orbits is added to those for the equatorial circular orbits.
We notice that the new curve seems to touch one of the equatorial ones.  This is indeed true. When $D\gamma v^3 = -\frac{2}{9}$, there is a bifurcation of critical points in which the equatorial saddle (green and beginning of orange) absorbs the pair of non-equatorial local minima (red) and becomes an equatorial minimum (rest of orange). From (\ref{eq:eqE},\ref{eq:eqL}), this happens at $E=\gamma+\tfrac{1}{3Dv}, L_z = \tfrac{1}{3v}$.

\begin{figure*}[t]
    \centering
    \includegraphics[width=\textwidth, height=\textwidth, keepaspectratio]{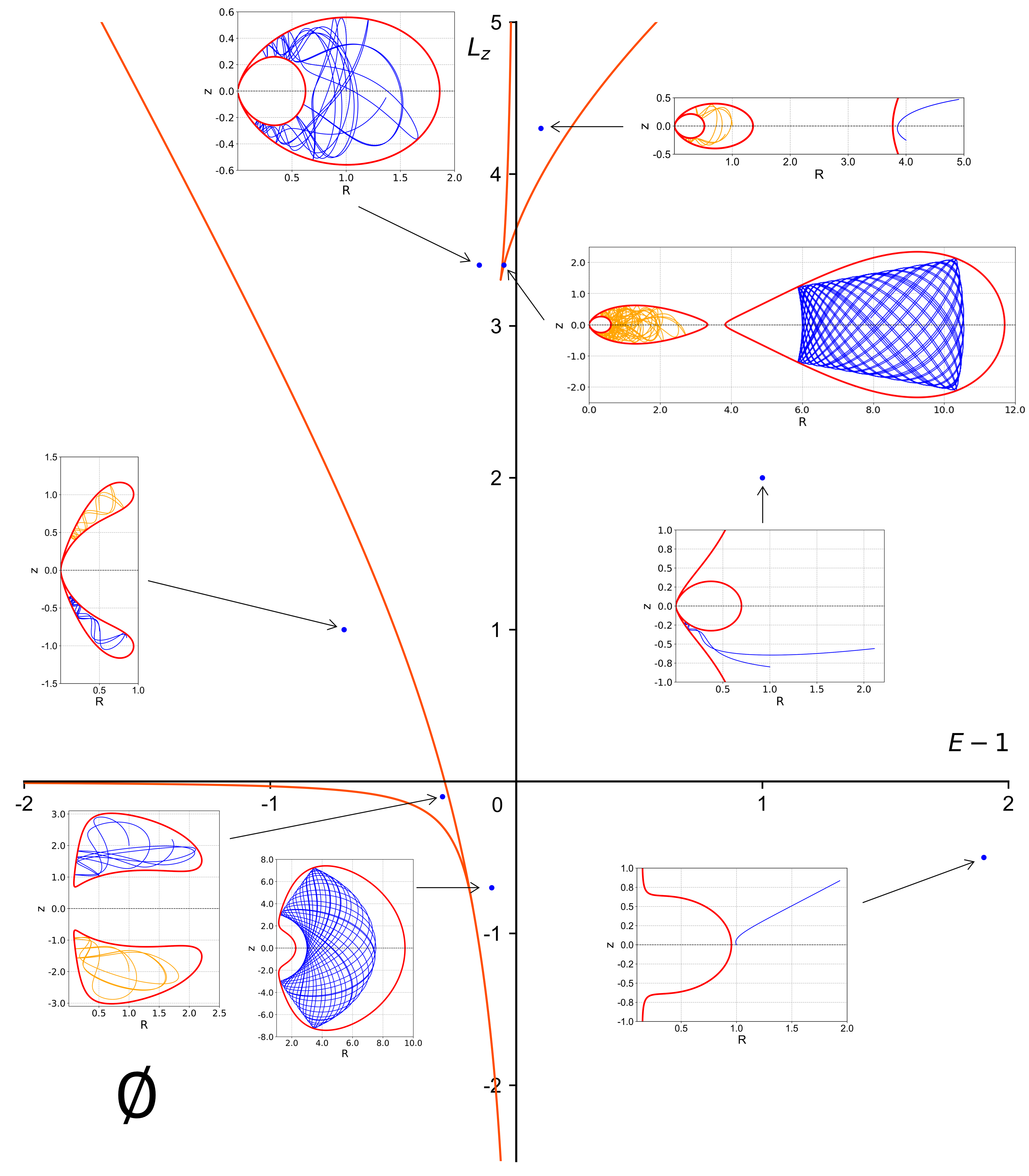}
    \caption{Shapes of Hill's region in the various domains in the space of $(E,L_z)$ for $D=3$ ($\emptyset$ means the Hill's region is empty), including the projections of some typical orbits at the indicated parameter values. }
    \label{fig:regionsa}
\end{figure*}

The conclusion is that the Hill's region has one of the forms given by the samples in Figure~\ref{fig:Hill} or is empty.  For $L_z>0$ there is a component with horns going into $r=0$.  For $L_z\le 0$ the Hill's region is bounded away from $R=0$ and may be empty.  For $E>1$ there is an unbounded component. The number and shapes of components are given by the parameter-domains in Figure~\ref{fig:regionsa}.

In this figure, we also plot the projections of some sample orbits.

\section{Poincar\'e plots}

For components of the Hill's region intersecting the equatorial plane, 
we can take $z=0, p_z>0$, as a Poincar\'e section $\Sigma$ (cf.~ref.~\onlinecite{BT}).   $\Sigma$ can be considered to be the set of $(R,p_R)$ where
$$E+\tfrac{1}{R} > \sqrt{1 + (\tfrac{L_z}{R}-\tfrac{D}{R^2})^2 + p_R^2},$$
with $z=0$ and $$p_z = \sqrt{(E+\tfrac{1}{R})^2 -1 - (\tfrac{L_z}{R}-\tfrac{D}{R^2})^2 - p_R^2}.$$
The boundary of $\Sigma$ (where $z=p_z=0$) is an orbit of the equatorial reduced system (section~\ref{sec:equator}).

Considering the first return map to $\Sigma$ simplifies representing and understanding the dynamics, at least of those orbits that intersect $\Sigma$.  
The section is transverse to the vector field, because $\dot{z} = p_z > 0$.  Let $\Sigma^+$ be the subset for which there is a first return in positive time.  Define $T(R,p_R)$ to be the time to the first return of the orbit started at $(R,p_R)$ on $\Sigma^+$ and $P(R,p_R)$ to be the point of the first return.  $P$ is called the {\em Poincar\'e map}.  By transversality of $\Sigma$ and smoothness of the flow, the set $\Sigma^+$ is an open subset of $\Sigma$ and $P$ is smooth on it.

\begin{figure}[htbp]
    \centering
    \begin{subfigure}[b]{\columnwidth}
        \centering
        \includegraphics[width=\columnwidth,height=4.3cm,keepaspectratio]{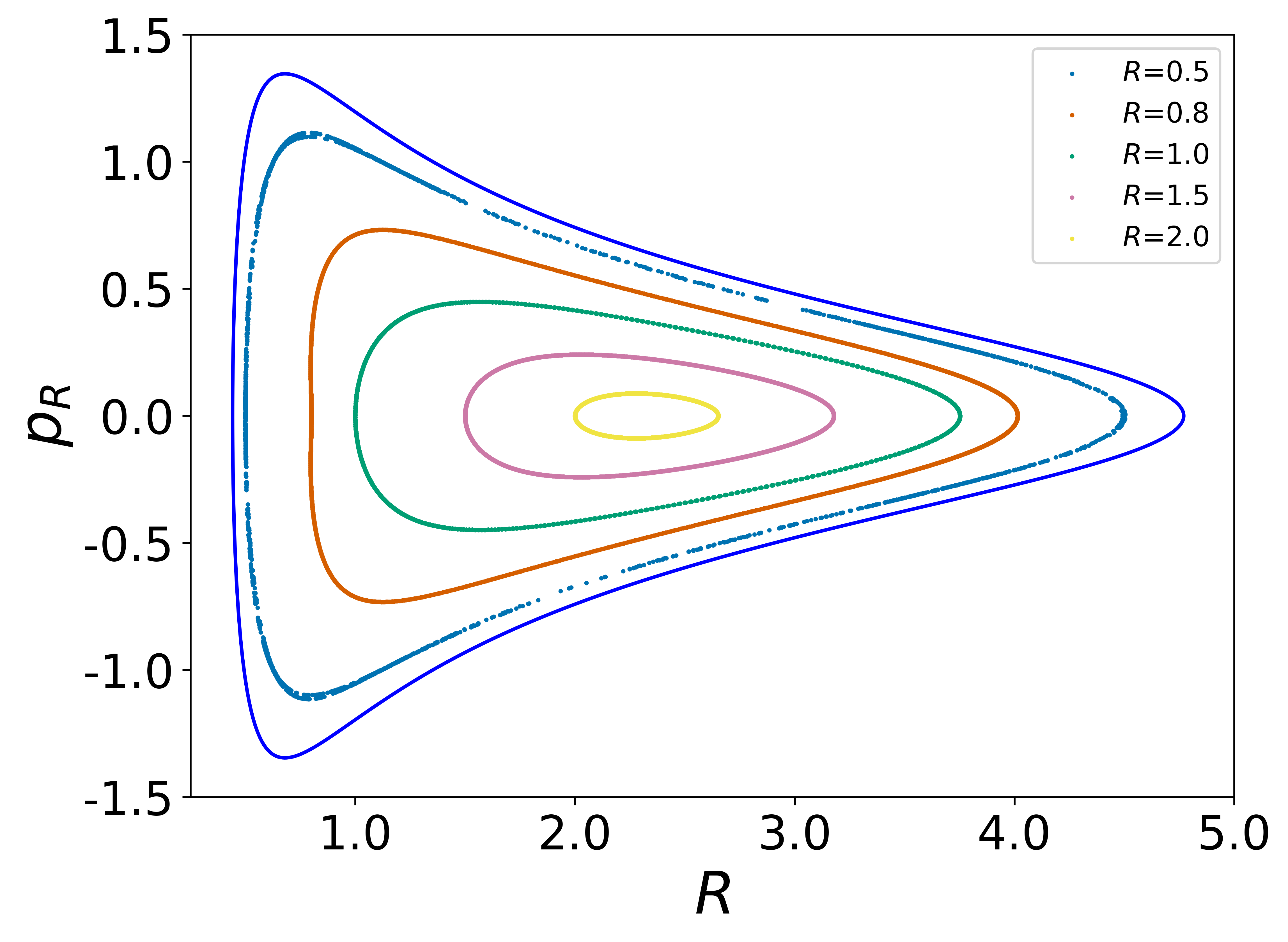}
        \caption{$(E, L_z, D)=(0.800, -0.600, 0.300)$}
    \end{subfigure}
    
    \vspace{0.3em}
    
    \begin{subfigure}[b]{\columnwidth}
        \centering
        \includegraphics[width=\columnwidth,height=4.3cm,keepaspectratio]{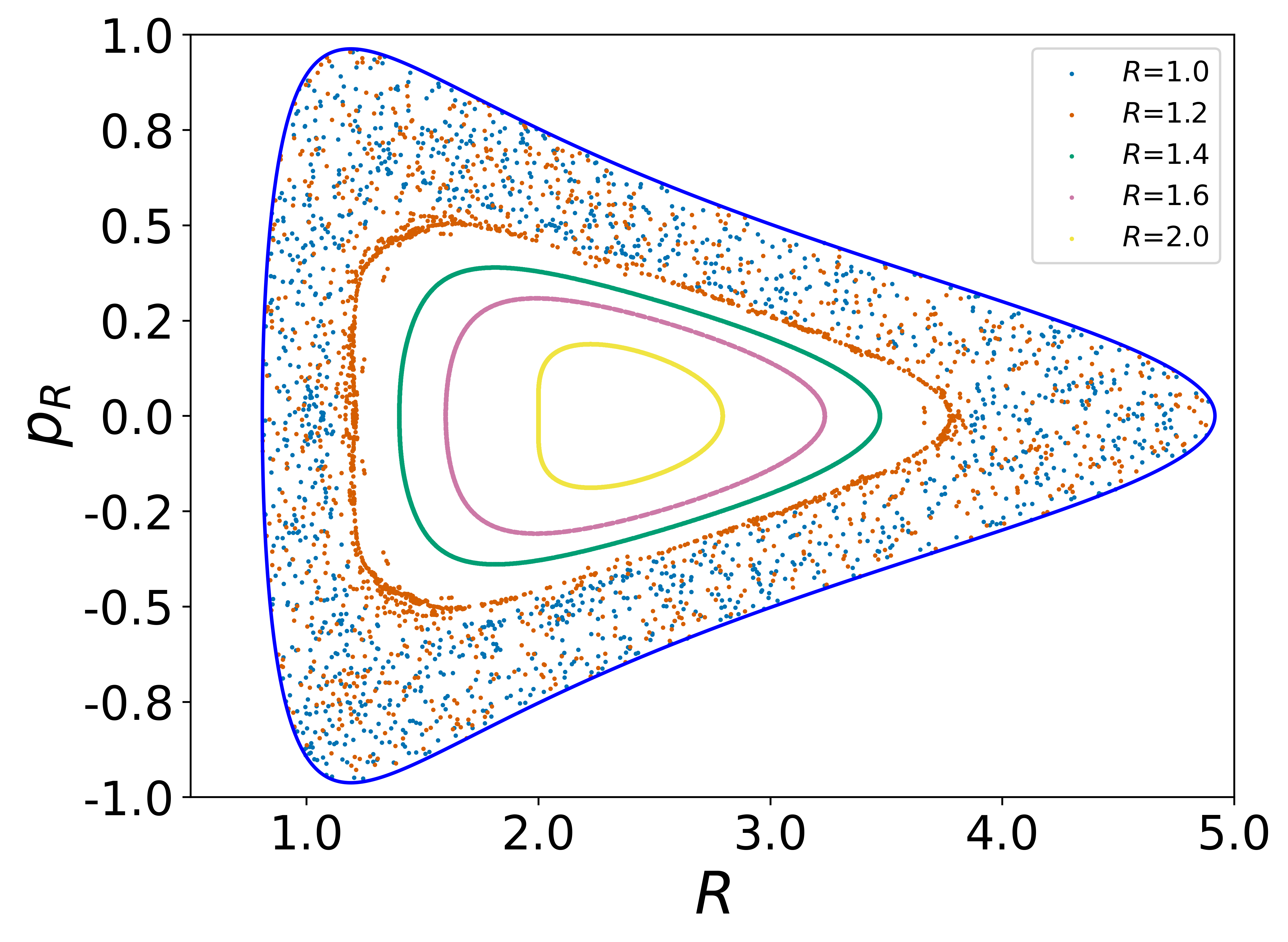}
        \caption{$(E, L_z, D)=(0.800, -0.200, 1.000)$}
    \end{subfigure}
    
    \vspace{0.3em}
    
    \begin{subfigure}[b]{\columnwidth}
        \centering
        \includegraphics[width=\columnwidth,height=4.3cm,keepaspectratio]{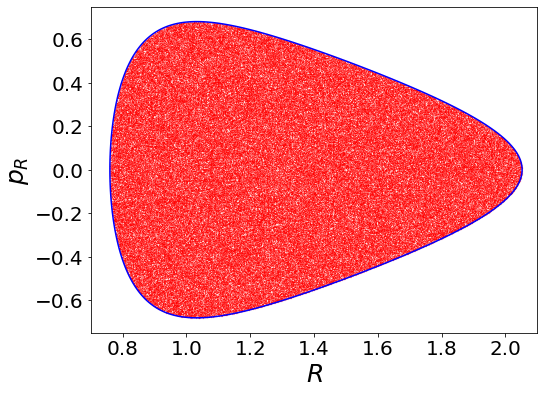}
        \caption{$(E, L_z, D)=(0.558, -0.283, 0.700)$}
    \end{subfigure}
    
    \vspace{0.3em}
    
    \begin{subfigure}[b]{\columnwidth}
        \centering
        \includegraphics[width=\columnwidth,height=4.3cm,keepaspectratio]{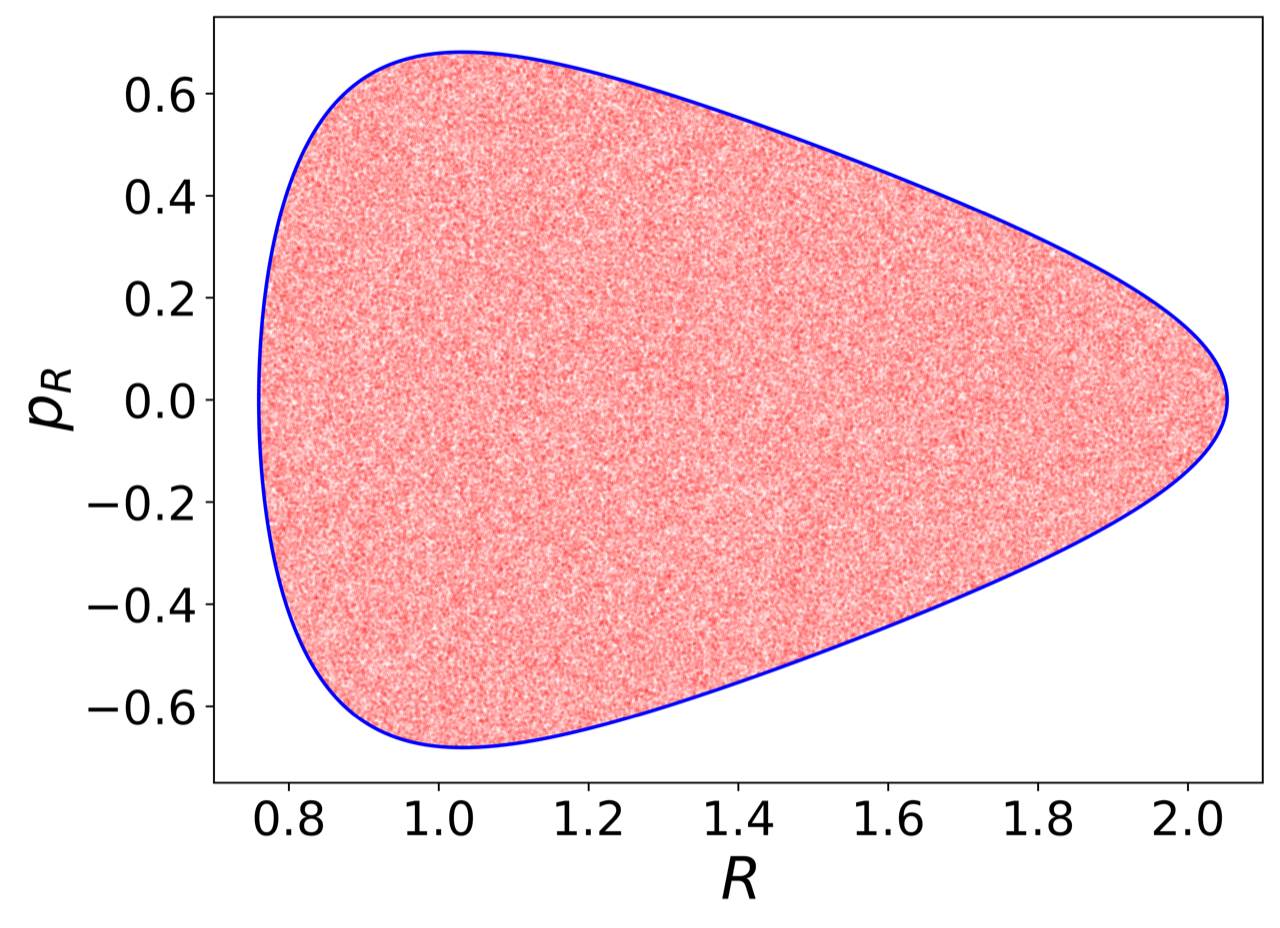}
        \caption{Transparency plot: $(E, L_z, D)=(0.558, -0.283, 0.700)$}
    \end{subfigure}
    
    \caption{Poincar\'e plots for selected values of $(E, L_z, D)$. The boundary of the Poincar\'e section (an equatorial periodic orbit in the reduced system) is shown in blue. We see that the behaviour varies from a) near-integrable, b) mixed, to c) near-perfect chaos.  In d) we replot the orbit from c) using an opacity of 0.1, showing that not only does the orbit appear to be nearly dense, it appears to have uniform density.}
    \label{fig:Pplots}
\end{figure}

Figure~\ref{fig:Pplots} shows some Poincar\'e plots (orbit segments for the Poincar\'e map) for selected values of $(E,L_z,D)$ and initial conditions.  We see a range of behaviours, on which we shall comment after discussion of the question of completeness of the section.

A transverse section is said to be {\em complete} if the forward and  backward trajectories of every point in the given Hill component crosses it (except for the bounding periodic orbit).  
A sufficient condition is that an ``angle'' variable evolves monotonically and without bound.
In particular, define angle $\alpha$ by 
\begin{equation}
\tan\alpha = p_z/z, 
\end{equation}
then use (\ref{eq:dotz}, \ref{eq:dotpz}) to obtain
$$\dot{\alpha} = \left(\frac{\dot{p}_z}{z}-\frac{p_z\dot{z}}{z^2}\right) \cos^2\alpha = -\left(\frac{3DRp_\phi}{r^5\gamma} + \frac{1}{r^3}\right)\cos^2\alpha - \frac{1}{\gamma} \sin^2\alpha,$$
which is negative if 
\begin{equation}
p_\phi > -\tfrac{\gamma r^2}{3DR}. 
\label{eq:condn}
\end{equation}
We will show that (\ref{eq:condn}) holds on the whole of the Hill component if $L_z$ is sufficiently negative (despite this looking counter-intuitive to the relation (\ref{eq:pphi}) and the requirement (\ref{eq:condn})).
For a bounded Hill component, if (\ref{eq:condn}) holds everywhere then $\dot{\alpha}$ is less than a negative constant on it.
Then from any point, both the forward and backward orbits cross $\alpha = \tfrac{\pi}{2}$ mod $2\pi$ at some times, which is our section.

\begin{figure}[htbp] 
   \centering
\includegraphics[height=2.0in]{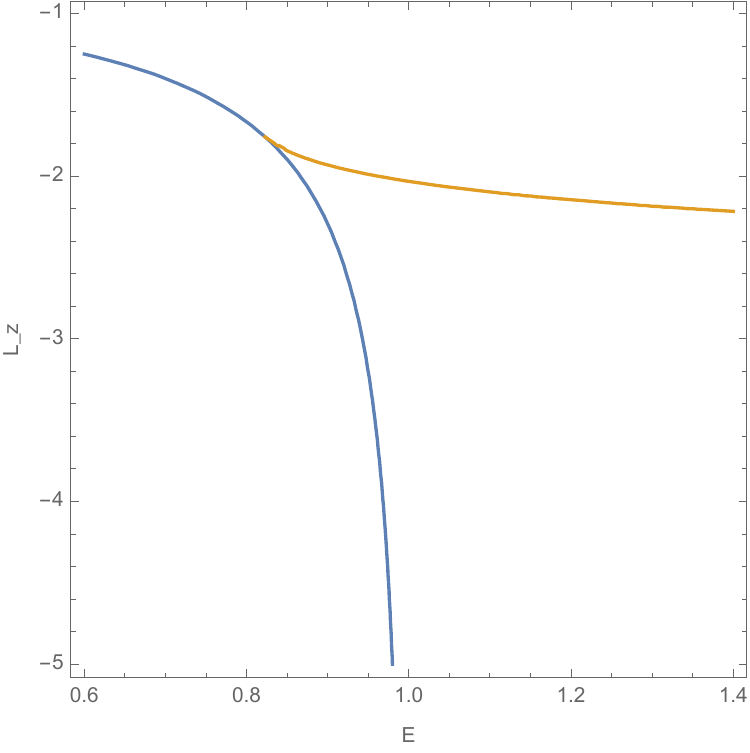} 
\caption{The region in $(E,L_z)$-space for which we prove that the section $z=0, p_z>0$ is complete is the cusped one between the red and blue curves. The region below the blue curve is where our bounds show that the Hill's region is empty, but from Figure~\ref{fig:circ} we know that the Hill's region is non-empty to the right of a curve asymptotic to $L_z = -\infty$ as $E$ increases to $1$.}
   \label{fig:bound}
\end{figure}

To obtain an explicit region of parameter space where there is a non-empty Hill component and (\ref{eq:condn}) holds on the whole of it, we will restrict attention to $L_z<0$.  The Hill's region can then be written as
\begin{equation}
1+ \tfrac{|L_z|^2}{R^2} + \tfrac{2|L_z|D}{r^3} + \tfrac{D^2R^2}{r^6} \le E^2 + \tfrac{2E}{r} + \tfrac{1}{r^2}.
\label{eq:ineq1}
\end{equation}
Using $1/R\ge 1/r = u$, writing $A = |L_z|^2-1$, and dropping the third and fourth terms from the lefthand side (which are positive and relatively negligible for large distance from the origin),
\begin{equation}
Au^2 - 2Eu + (1-E^2) \le 0 
\label{eq:ineq2}
\end{equation}
in the Hill's region.
We restrict further to $L_z < -1$, so $A>0$.
It follows that for $E \ge \sqrt{\frac{A}{1+A}}$,
\begin{equation}
u \le u_+ = \frac{E + \sqrt{E^2 - A(1-E^2)}}{A}
\end{equation}
(and if $E < \sqrt{\frac{A}{1+A}}$ then the Hill's region is empty).
Now
$$\frac{|p_\phi|}{\gamma}\frac{3DR}{r^2} = 3D\frac{|L_z|+DR^2/r^3}{r^2(E+1/r)} \le 3D \frac{|L_z|+D/r}{r(Er+1)}.$$
The latter expression is decreasing in $r$, so we can insert $r\ge 1/u_+$ to obtain
$$\frac{|p_\phi|}{\gamma}\frac{3DR}{r^2} \le 3D \frac{(|L_z|+D u_+)u_+^2} {(E+u_+)}.$$
This is less than 1 (and so condition (\ref{eq:condn}) holds) if
$$3D(|L_z|+Du_+)u_+^2 < E+u_+ .$$
We plot this (and the associated curve for empty Hill's region) in Figure~\ref{fig:bound} for $D=1$ as the region between the red and blue curves.  We see that the region in which we have established completeness of the section includes all cases of non-empty bounded Hill region for $L_z$ sufficiently negative, as claimed.
This can be proved for any value of $D>0$ but we don't give the estimates here.

We expect the true domain for completeness of the section to go up to the bifurcation point in Figure~\ref{fig:critpts}, because near the part of the curve where the critical point is elliptic, the Hill's region is small and the linearisation about the bounding periodic orbit has non-zero rotation.  Perhaps our estimates could be improved to show this (notably by throwing away less of the higher order terms in the step from (\ref{eq:ineq1}) to (\ref{eq:ineq2}); but the estimate $R\le r$ is close to optimal for small Hill's region).

On the other hand, for $L_z$ not sufficiently negative, the bounding periodic orbit has linearised rotation rate zero around it and is hyperbolic (because associated with a hyperbolic equatorial circular orbit), so the forward orbits of points on its local stable manifold do not intersect the section (and the same for backwards orbits of points on its local unstable manifold).  Indeed, it can be expected that there is a set of positive volume whose orbits do not intersect the section, so the Poincar\'e map would not tell us anything about them.

The boundary in $(E,L_z)$ space between parameter values for which the equatorial periodic orbit has zero or non-zero linearised rotation rate can in principle be determined by solving a Hill's equation for the Floquet multipliers (the linearised equation in $(z,p_z)$ about the periodic orbit $R(t)$; compare the curve for stabilisation of an inverted pendulum by oscillation of its support), but we have not attempted that.

If the component $C$ of the energy-surface corresponding to the chosen Hill component has finite volume $V_C$ then $\Sigma^+$ is a subset of full measure in $\Sigma$, by volume-preservation of the flow.  So $\Sigma \setminus \Sigma^+$ is negligible in a measure-theoretic sense.
For components of Hill's region that are bounded away from $r=0$ and $\infty$, the volume is finite.  This is because energy-surface volume $\mu$ is defined on each energy-surface by 
$$\tfrac12 \omega \wedge \omega = \mu \wedge dH,$$ 
so $V_C = \int_C \mu$.  The easiest way to bound it for our problem is to express it as 
$$V_C = \int_C 2\pi \sqrt{K}\, dR dz$$ 
over the Hill component, because for each point of the Hill component there is a circle of possible momenta, with magnitude $\sqrt{K}$, where 
\begin{equation}
K = \left(E+\tfrac{1}{r}\right)^2 -1 - \left(\tfrac{L_z}{R}-\tfrac{DR}{r^3}\right)^2.
\end{equation}
Under the stated conditions, this integral is finite.  Thus the return map is defined on a set of full measure in $\Sigma$, which is consistent with the figures.

For components with horns going to $r=0$, the volume is also finite but to prove it requires a little more work.  The centre curve of the horn is $L_z r^3= DR^2$.  We can write this as $R=R_c(z)$ with $R_c(z)^2 \sim \frac{L_z}{D} z^3$.  Near the centre curve, 
$$\tfrac{L_z}{R}-\tfrac{DR}{r^3} \sim -\tfrac{2D}{z^3}(R-R_c(z)).$$  Also, $E+\frac{1}{r} \sim \frac{1}{z}$.  So the horn is given asymptotically by $R = R_c(z) \pm \Delta R(z)$ with $\Delta R \sim \frac{z^2}{D}$.  Then the integral from the part of the horn with $z \le \zeta$ is asymptotically $$\int_0^\zeta \tfrac{z}{D} dz = \tfrac{\zeta^2}{2D} < \infty.$$  The contribution from the rest of the Hill component is bounded.

\begin{figure}[htbp] 
   \centering
\includegraphics[height=2.2in]{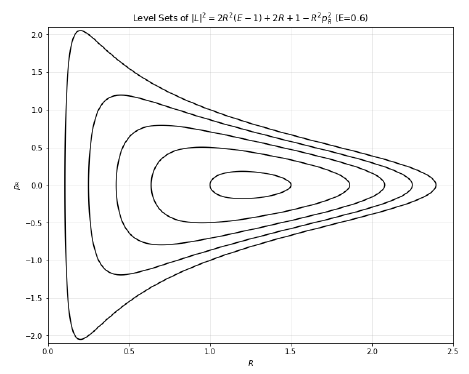}
\caption{Level sets of $L^2$ in the Poincar\'e section $z=0$.}
   \label{fig:PD=0}
\end{figure}

For $D=0$ we find (not shown) that the Poincar\'e plots look integrable.  Indeed, it is easy to see an additional constant of the motion, namely the modulus-squared of the angular momentum vector, 
$$L^2 = r^2 p^2 - (\bm{r} \cdot \bm{p})^2.$$ 
This can be evaluated to  
$$L^2 = (zp_R-Rp_z)^2 + (1+\tfrac{z^2}{R^2}) L_z^2.$$
Then a little calculation confirms that $\frac{d}{dt}L^2 = 0$.  Some level sets of $L^2$ are plotted in Figure~\ref{fig:PD=0}.

When $D \ne 0$ but small, Poincar\'e plots like Figure~\ref{fig:Pplots}(a) suggest that the motion is close to integrable.  This is plausible because for $D=0$ the motion is integrable with additional constant $L^2$ and the frequency of precession varies non-trivially, and for $DR \ll r^3$ the perturbation counts as small, so KAM theory applies.  To complete the argument is somewhat delicate, however, because the precession-frequency goes to zero in the non-relativistic limit.  
Another regime in which near-integrable behaviour can be expected is when the Hill's region is a small disk crossing $z=0$; in this case the motion is close to separable into $R$ and $z$ in the formulation (\ref{eq:J}).
We leave analysing these situations for future work.

For larger $D$, 
the Poincar\'e plots suggest strongly that the motion is not integrable.  That is in interesting contrast to geodesic motion in the Kerr metric, to which Aubry's system can be considered as an approximation.  As already mentioned, Carter found a third constant of the motion (or fourth in space-time) for the Kerr metric, rendering its geodesic flow integrable \cite{Ca}.  The corresponding continuous symmetry is the Hamiltonian flow of the Carter constant (for the proper-time dynamics in space-time).  The Carter constant is quadratic in the 4-velocity (like the Hamiltonian for the dynamics).  A Carter constant is in fact a general feature of Einstein vacuum space-times with a rotation symmetry \cite{WP}.  

Not only does the motion for $D>0$ appear to be non-integrable, for some values of parameters the Poincar\'e section appears to be filled densely by a single orbit, e.g.~Figure~\ref{fig:Pplots}(c).  Even more, the chosen orbit appears to fill out the section with uniform density.  This is remarkable.  To confirm this, we replotted in Figure~\ref{fig:Pplots}(d) with opacity 0.1, so that any variations in density after pixelisation could become apparent, and none are.
  
Sometimes, purely topological aspects of a system can force chaotic dynamics \cite{M01}, but not necessarily on a set of positive measure, let alone full measure and with unifrom density.  Note that special examples, called pseudo-Anosov, can be constructed in which this topological chaos has full measure (and typical orbits have uniform density); there is a chance something close to this could occur in the case that the bounding orbit of the section has a non-zero half-integer rotation-rate (making ``1-prongs'' on the boundary), but it likely to require special conditions.  The case of zero rotation-rate is more relevant, however.  It can not be pseudo-Anosov, but we give a sketch of a suggested explanation for the observed near-perfect chaos in the next section.

Note that some orbits in the component of the Hill's region might not intersect the section, so the Poincar\'e map might not give the complete dynamics, even measure-theoretically.  The accessible volume from the Poincar\'e section is given by 
$$V_{acc} = \int_{\Sigma} T(R,p_R)\ dR dp_R.$$
This simple formula is a consequence of preservation of energy-surface volume $\mu$, as defined above.  The proof is that the flux of energy-surface volume  $i_X \mu = \omega$, so the volume swept out by unit area on the section in time $T$ is $T\omega$.

If $V_{acc}$ is less than the volume $V_C$ of the component $C$ then a positive volume of orbits is missed by the Poincar\'e map.
There are some parameter regimes where one could probably prove this.  In particular, when the Hill's region has a small neck on $z=0$ and large bulges for $z \ne 0$ it is plausible that $V_{acc} \to 0$ as the neck width goes to $0$ but $V_C$ remains bounded away from $0$.  The only way to avoid this would be if the return time function $T$ diverges sufficiently fast.  This scenario will be discussed in the next section.

One case in which $V_{acc} < V_C$ is if the boundary of a Poincar\'e section is a hyperbolic periodic orbit and its stable and unstable manifolds join up.  Then the invariant manifolds separate the space into three components, one of which is the accessible volume from the section and the other two are invariant volumes in $z > 0$ and $z<0$ respectively. 
This will be sketched in Figure~\ref{fig:integ}.

Two questions arise for future work:~can one modify the construction of the Poincar\'e map to capture almost all of the dynamics?  And can one make a Poincar\'e section for components of the Hill's region that do not intersect the equatorial plane?

On the first question, a general answer is suggested in ref.~\onlinecite{Du}.  In our context it could probably be better achieved by adding  transverse sections spanned by two index 1 ``brake'' periodic orbits to be defined in the next paragraph and described in the next section.

On the second question, any Hill's component homeomorphic to a disk contains a minimising brake periodic orbit \cite{S} (use the formulation (\ref{eq:J}) to apply Seifert's result directly).  A {\em brake orbit} of a reversible Hamiltonian system is an orbit whose projection $\Pi$ to $(R,z)$ connects two points of the boundary. 
By reversibility it then reverses along the same path (but with the opposite velocity).  In the $z$-symmetric Hill components, there is an obvious brake orbit along $z=0$, but Seifert's construction generalises this to non-symmetric contexts.
Then one can use the brake orbit to make a Poincar\'e section.  Choose an orientation for $\Pi$.  Let $\Sigma$ be the set of $(R,z, p_R, p_z)$ such that $(R,z) \in \Pi$, $$p_R^2+p_z^2 = \left(E+\tfrac{1}{r}\right)^2 - 1 - \left(\tfrac{L_z}{R}-\tfrac{DR}{r^3}\right)^2,$$ 
and such that the oriented angle from the tangent to $\Pi$ to $(p_R,p_Z)$ is in $(0,\pi)$.  Then $\Sigma$ is transverse to the vector field and the same considerations as for equatorial sections apply.  To use this would require constructing a brake orbit numerically (which is not hard; we do it for the first question in the next section), but would still leave open the risk that not every orbit crosses it.

\section{From integrable to near-perfect chaos}

\begin{figure}
    \centering
    \includegraphics[height=2in]{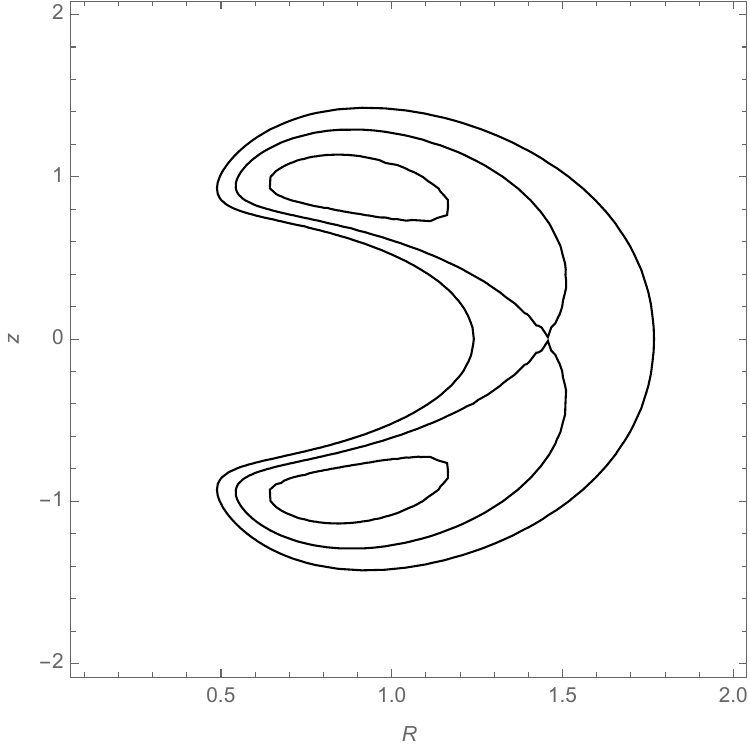}
    \caption{Example of Hill's component containing an equatorial saddle and non-equatorial minima; $D=1, L_z=-0.348, E=0.56$.}
    \label{fig:bean}
\end{figure}

Here we sketch a suggested explanation for how near-perfect chaos can arise in a Hill's component that is a $z$-symmetric topological disk containing a saddle of $J$ on $z=0$ and a pair of minima of $J$ with $z \ne 0$, such as in Figure~\ref{fig:bean}.  With slight modification the same argument is expected to apply to the case of $z$-symmetric horned Hill's regions crossing $z=0$.

The scenario has large potential applicability outside this problem too, in particular the $z$-symmetry is not required.  The basic context is a double-well system of 2 DoF.  Aspects of this were understood in the classical theory of chemical reactions years ago\cite{DL,O+}, under the name "reactive islands" (for a recent article along this line, giving more references, see ref.~\onlinecite{NKW}), but we feel that the geometry of the energy-level has not been properly described.  Another context in which it arises is celestial mechanics, for example transition of test particles over the Lagrange points in the circular restricted three-body problem\cite{MR,D+}.  It occurs in models of ship capsize too\cite{NR} and the double pendulum\cite{KBKB}.
One notable future application would be to magnetic confinement of charged particles in a finite sequence of magnetic bottles, following ref.~\onlinecite{BMN}.  This is the basis for ``quasi-isodynamic'' stellarators, such as Wendelstein-7X.

\begin{figure}[tbp] 
   \centering
\includegraphics[width=3.2in]{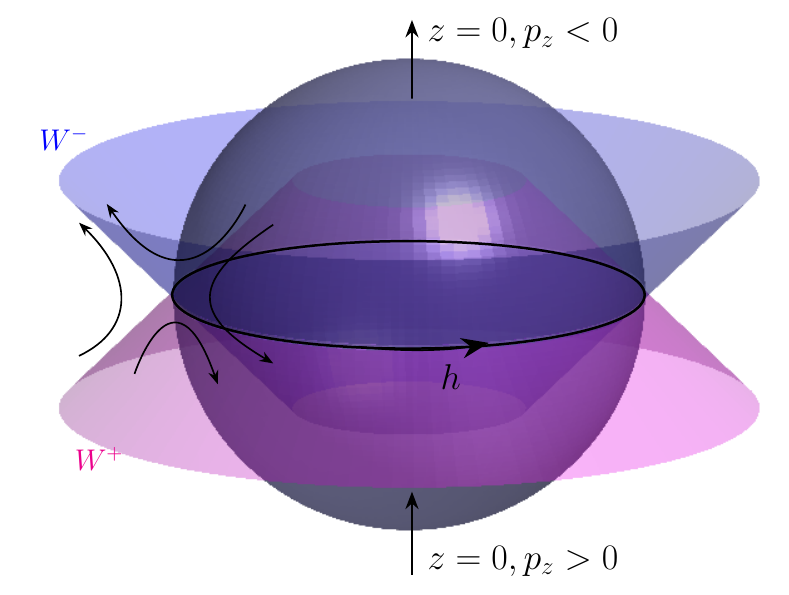} 
   \caption{The local contracting submanifolds of the hyperbolic periodic orbit $h$ relative to the section $z=0$. $z>0$ inside the indicated sphere and $z<0$ outside it.
}
   \label{fig:hyppo}
\end{figure}

The energy-level is a 3-sphere.  The subset with $z=0$ cuts it into two 3-balls, corresponding to $z>0$ and $z<0$. The subset with $z=0$ is a 2-sphere.  It contains a hyperbolic periodic orbit $h$ on which $p_z=0$.  The hyperbolic periodic orbit cuts the 2-sphere into two hemispheres, corresponding to $p_z>0$ and $p_z<0$.  The one with $p_z>0$ is what we chose as Poincar\'e section.  We call the other one the ``dual section''.
The local forwards and backwards contracting submanifolds of $h$ look as in Figure~\ref{fig:hyppo} relative to the 2-sphere (compare ref.~\onlinecite{M91}).  
It is possible to draw the 2-sphere as the horizontal plane, with one hemisphere being the disk inside $h$ and the other being its complement plus the point at infinity, but we find it less convenient.  Nevertheless, when it comes to considering Poincar\'e maps, we might want to flatten the hemispheres in that way.

\begin{figure}[htbp] 
   \centering
\includegraphics[width=2in]{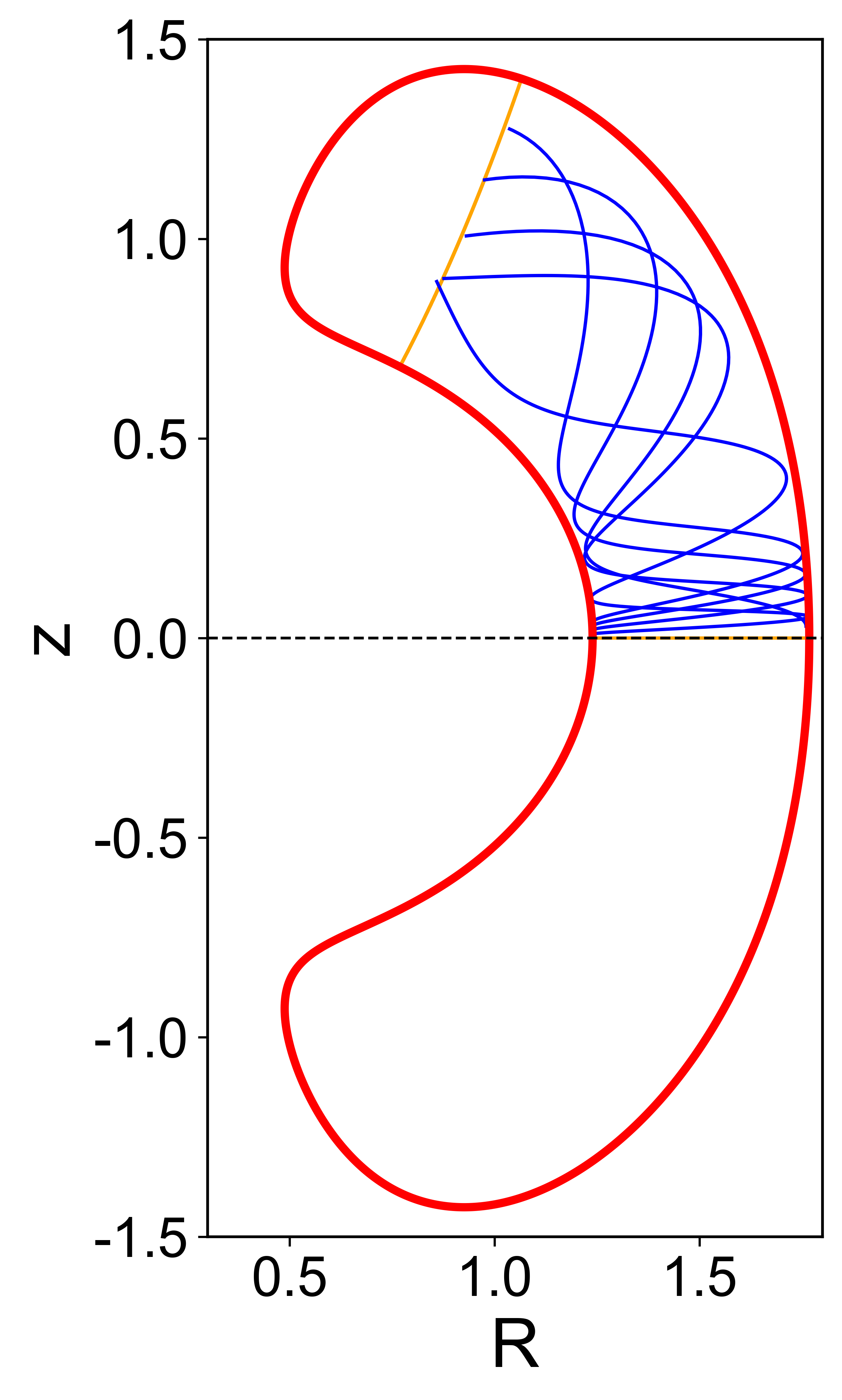} 
   \caption{The Hill's region for $D=1, L_z=-0.348, E=0.56$, showing the hyperbolic (on $z=0$) and index 1 (in the top half) brake periodic orbits and some orbits on one side of the unstable manifold of $h$ up to their first crossing with $e$. }
   \label{fig:HillwWu}
\end{figure}

Focus on the dynamics inside the 3-ball for $z>0$ (by $z$-symmetry, the 3-ball for $z<0$ is equivalent).  There is a periodic orbit $e$ of Poincar\'e index $+1$ in $z>0$, which bounces between two points on the boundary of the Hill's component (a brake orbit, which we believe can be obtained by a similar argument to Seifert; there is a corresponding one for $z <0$). We call it $e$ because we think of the case where it is elliptic, though in some parameter regimes it is inversion hyperbolic (in particular, this turns out to be the case in Figure~\ref{fig:HillwWu}).
It can be found numerically by shooting from the boundary of the Hill's component to a local section $p_R=0$ near the opposite side of the boundary and varying the initial point to make $p_z=0$ on the section.
The local backwards contracting submanifold of $h$ is a cylinder whose projection crosses $e$, see Figure~\ref{fig:HillwWu} (cf.~pictures in ref.~\onlinecite{O+,MR,NR}, and in ref.~\onlinecite{BMN} for an analogous case in a magnetic mirror machine, though none of these include the index 1 orbit). 

In the energy level, $h$ can be spanned by a disk that contains $e$ and is transverse to the flow except on $h$ and $e$.  To construct such a disk numerically, one can choose a smooth function $F(R,z)$ (of the form $z g(R,z)$) that is $0$ on $h$, $1$ on $e$ and has nowhere-zero derivative.  Then a suitable section is the subset of the energy level where $\dot{F}=0$. For transversality off $h$ and $e$ we ask for $\ddot{F} \ne 0$ there.
We draw this disk horizontally in the next energy-level pictures.  The part inside $e$ has $\ddot{F}<0$; the part between $h$ and $e$ has $\ddot{F}>0$.

\begin{figure}[tbp] 
   \centering
\includegraphics[width=3.2in]{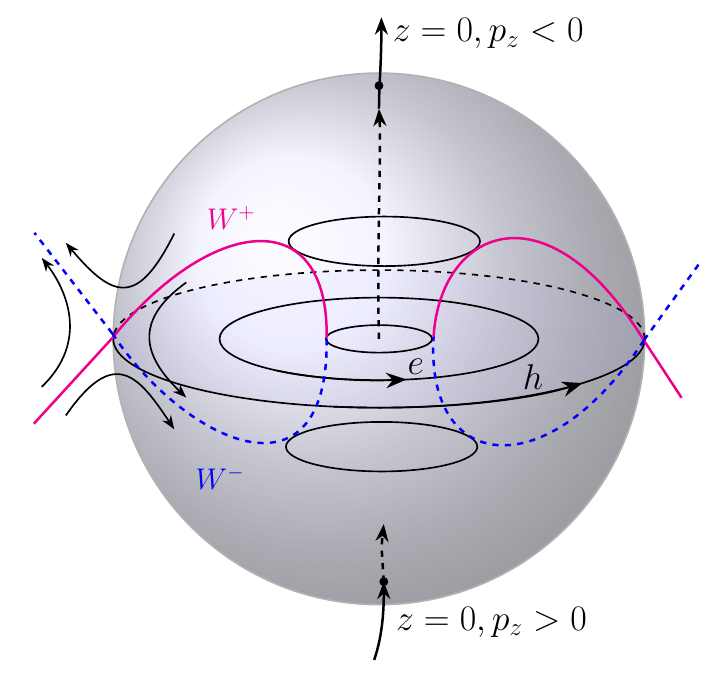} 
\caption{The ball $z\ge 0$ in the integrable case.
}
   \label{fig:integ}
\end{figure}

If the dynamics has a rotation symmetry (not necessarily from isometry of configuration space; it can be from separability of vertical and horizontal motion or the adiabatic invariant to be described in sec.~\ref{sec:adinv}) that leaves $h$ and $e$ invariant then the system is integrable and the phase portrait in $z>0$ looks like Figure~\ref{fig:integ}, where the inner branches of $W_\pm$ coincide.  
The map from $z=0, p_z>0$ to $z=0, p_z<0$ preserves the circles of rotation, the map on each circle being a rotation depending on the rotational part of the dynamics and the time it takes (which goes to infinity at the boundary $h$).  The return map to $z=0, p_z>0$ is the composition of this map with the corresponding one for $z<0$.  By $z$-symmetry, the second map is equivalent to the first, so it is enough to consider just the first.  
Note that there is a significant volume that is not accessible from the Poincar\'e section, namely that enclosed by the separatrix, including $e$.  The dynamics in it is a ``vortex'' around $e$.

\begin{figure}[htbp] 
   \centering
\includegraphics[width=3.2in]{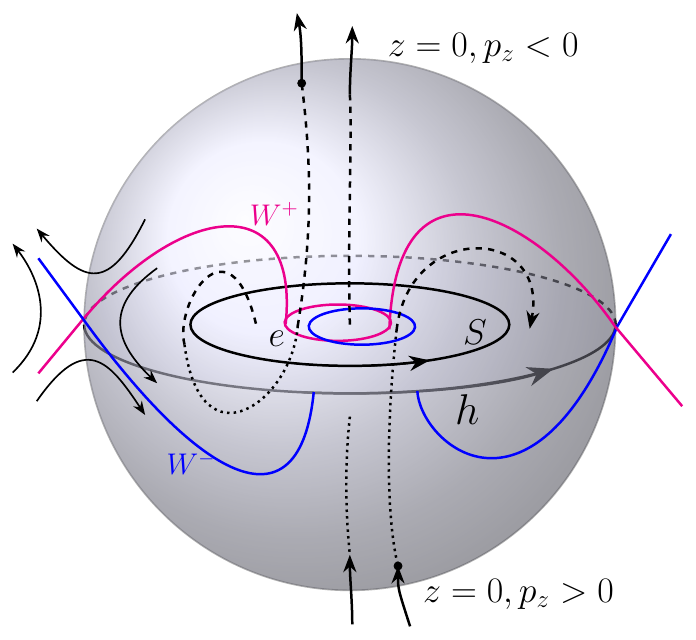} 
\caption{The ball $z \ge 0$ in a near-integrable case. 
}
   \label{fig:nearinteg}
\end{figure}

For small deviation from integrable, the picture deforms to that of Figure~\ref{fig:nearinteg} (compare ref.~\onlinecite{BMN}).  
There are two (or more) lobes in the horizontal plane, through which orbits make transitions between ``free'' and ``trapped'' motion.  By ``trapped'' we mean that it circulates around $e$ and can not reach the Poincar\'e section again until it makes a transition to ``free''.  The dynamics in this ``vortex'' is in general more complicated than in the integrable case. 
The accessible region is bounded by a hierarchy of invariant tori and hyperbolic invariant sets around $e$.

The map from Poincar\'e to dual section can be considered to be the composition of three maps:~from Poincar\'e section to the disk $IN$ in a section $S$ spanning $e$ (part of the previously mentioned section spanning $h$ and containing $e$), from $IN$ to $OUT$ by iterating the return map $f$ to $S$, and from $OUT$ to the dual section.

We start by analysing the map $f$. It sends a foliation of the annulus between $e$ and $OUT$ by closed curves around $OUT$ to a foliation of the annulus between $e$ and $IN$ by closed curves around $IN$, with a rotation rate that goes to infinity logarithmically as the boundary of $OUT$ is approached, because of the time spent near $h$.

\begin{figure}[htbp] 
   \centering
\includegraphics[width=3in]{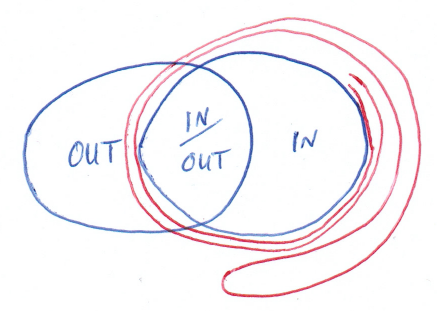} 
\caption{Sketch of the image of the $IN$ lobe under $f$. }
   \label{fig:image}
\end{figure}

Thus the image under $f$ of the part of $IN$ that doesn't go straight out is infinitely wrapped around the boundary of $IN$, as in Figure~\ref{fig:image}.
The intersection of $f(IN)$ with $OUT$ is an infinite sequence of strips in $OUT$ (there could also be a ``nose'').  They come from an infinite sequence of similar strips in $IN$, but $f$ interchanges the short and the long sides, so it is highly hyperbolic. 
Successive images of the remainder of $IN$ get wrapped around $IN$, producing further infinite sequences of strips in $OUT$ (and possible further noses).  All but a set of measure zero of $OUT$ is covered as iteration goes to infinity. 

\begin{figure}[htbp] 
   \centering
\includegraphics[width=3.2in]{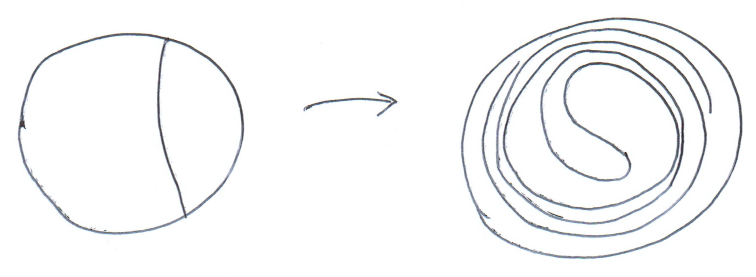} 
\caption{Sketch of the map from a chord in the exit disk to a ying-yang curve in the dual section. }
   \label{fig:yinyang}
\end{figure}

The shapes of the map from the Poincar\'e section to the entry disk, and that from the exit disk to the dual section, are not hard to analyse.  Focussing on the former, it consists principally in a rotation that goes to infinity at the boundary, because the time taken from the section to the entry disk goes to infinity like a logarithm of the (reciprocal) distance to the boundary and during that time, the orbit rotates roughly the same as $h$ does.  This implies that the preimage of a chord in the entry disk (such as the boundaries of the previously derived strips) is an infinite ``yin-yang'' curve in the Poincar\'e section.  Similarly, the image of a chord in the exit disk is an infinite yin-yang curve in the dual section, as shown in Figure~\ref{fig:yinyang}.  A sequence of non-intersecting chords produces a nested sequence of infinite yin-yang curves.  

\begin{figure}[tbp] 
   \centering
\includegraphics[width=3.1in]{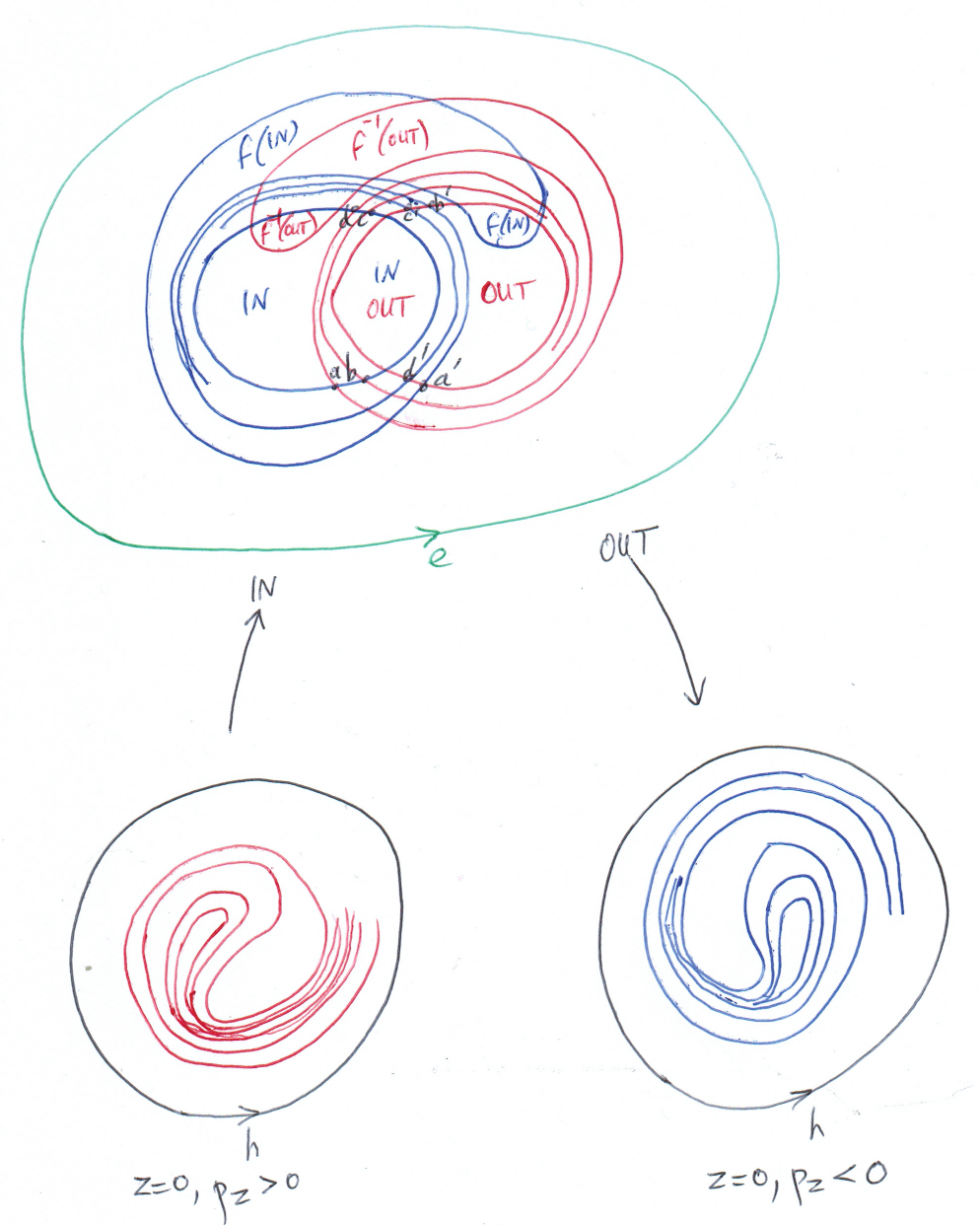} 
\caption{Sketch of part of the transit map from Poincar\'e section to dual section. }
   \label{fig:transit}
\end{figure}

Thus the map from Poincar\'e section to dual section takes one nested set of infinite yin-yang curves to another one.  But they approach the boundary in opposite directions.  More importantly, $f$ interchanges short and long sides.
Figure~\ref{fig:transit} shows a bit of it (only the first iterate of $f$ has been taken into account and only the first wrapping around).

\begin{figure}[tbp] 
   \centering
\includegraphics[width=3.1in]{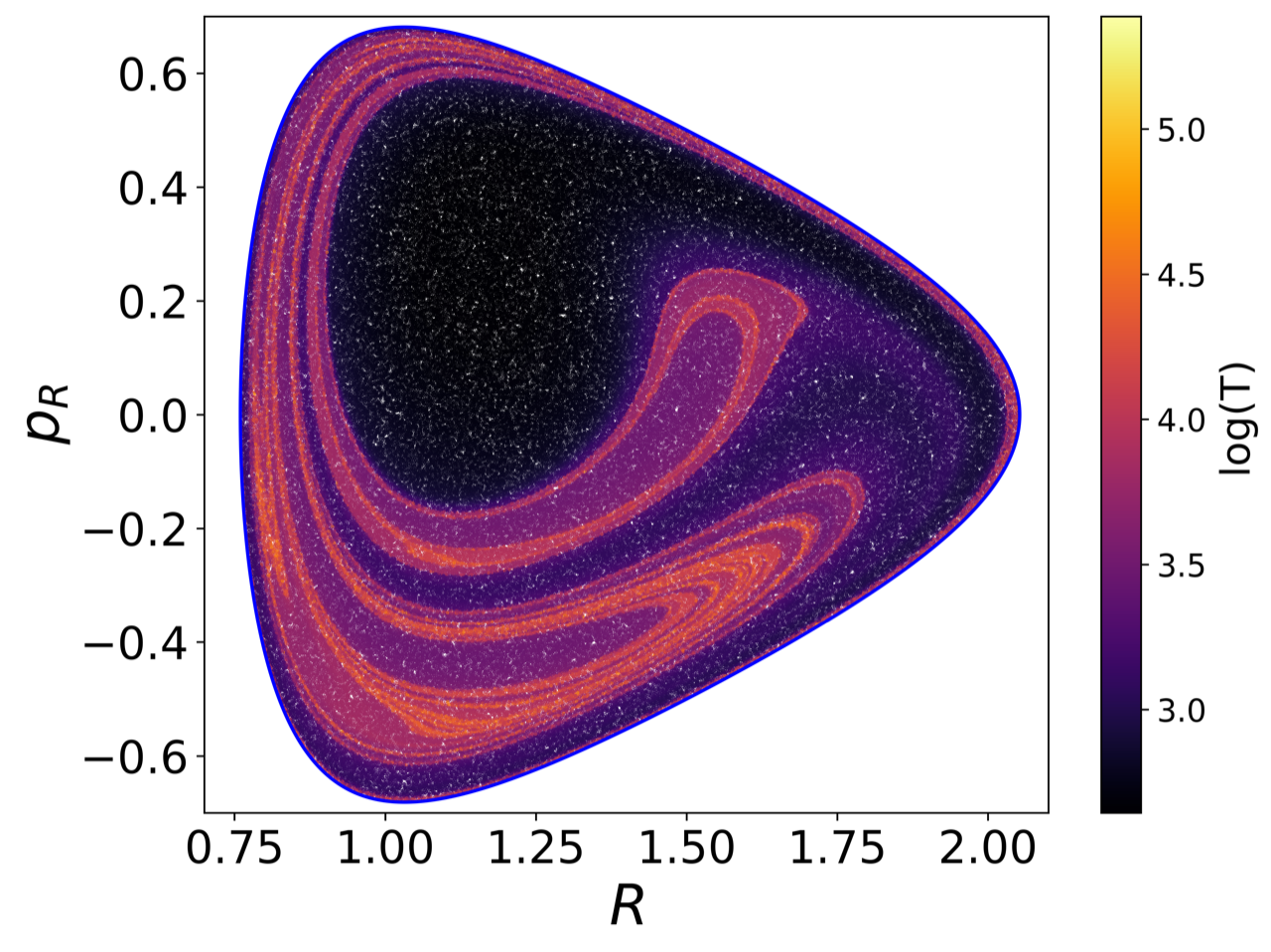} 
\includegraphics[width=3.1in]{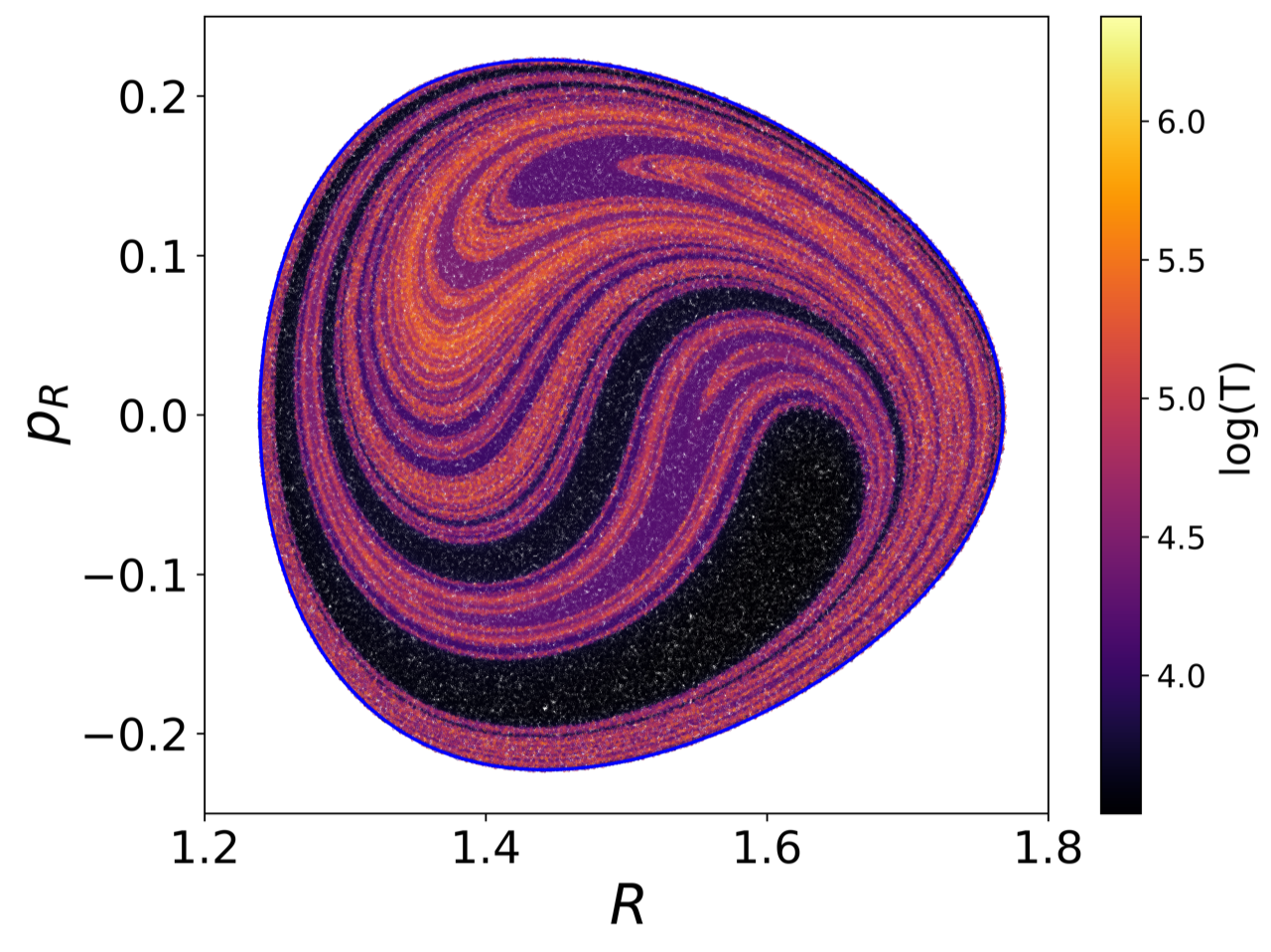}
\caption{Poincar\'e plots for (upper) $D=0.7$, $L_z = -0.283$, $E=0.558$, (lower) $D=1$, $L_z = -0.348$, $E=0.56$, with points coloured according to return time $T$.}
   \label{fig:PsecReturnTime}
\end{figure}

For comparison, Figure~\ref{fig:PsecReturnTime} shows the Poincar\'e plot with points coloured by the return time.  The yin-yang curves are clearly visible.

\begin{figure}[htb] 
   \centering
\includegraphics[width=2.3in]{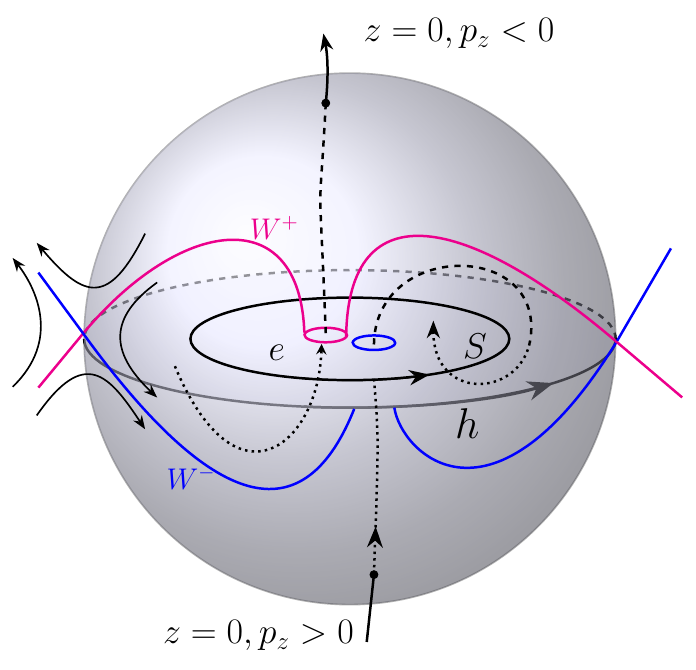} 
\caption{The ball $z \ge 0$ in a far from integrable case. 
}
   \label{fig:farfrominteg}
\end{figure}

For larger deviation from integrability, the contracting submanifolds of $h$ might miss each other on their first transit through the horizontal plane, as in Figure~\ref{fig:farfrominteg} (compare ref.~\onlinecite{BMN} again).
Then every orbit from the Poincar\'e section gets trapped.  It goes round the ``vortex'' for some time until  with probability 1 it eventually exits and reaches the dual section.  The picture is similar to the previous case, except there is no $IN/OUT$ region.
By conservation of volume and finiteness of the trapped volume, the entry disk is partitioned, up to a set of zero area, into pieces corresponding to different numbers of revolutions in the vortex before reaching the exit disk in a similar set of pieces.  This could be quite complicated because of the dynamics in the vortex, but most of it consists of slicing the disk by an infinite sequence of disjoint chords.  Then the same discussion as above applies.

The result is probably best plotted and analysed in a coordinate system with the logarithm of reciprocal distance to the boundary as radius.  
It would be good to flesh out this sketch analysis more thoroughly. 
The closest we have seen in the literature is ref.~\onlinecite{KLMR}.  One route could be to study the dynamics close to the curve of circular equatorial orbits in parameter space.  Then the ``neck'' is small and it might suffice to follow the (1D) stable and unstable manifolds of the critical point.

We emphasise that this near-perfect chaos is only on the section spanning the hyperbolic orbit.  It is expected to coexist with near-integrable behaviour on the sections spanning the index 1 orbits when elliptic.

One might ask how reliable are long-time numerics for evidence of near-perfect chaos.  The theory of shadowing allows one in principle to determine whether there is a true trajectory within a small distance of a numerical one; see Ref.~\onlinecite{HJ} and the many preceding references it cites.  We did not implement this, but based on the accumulated experience of such studies, we are confident that our numerics do represent genuine behaviour.

\section{Adiabatic invariant}
\label{sec:adinv}

We have shown already in Figure~\ref{fig:orbits}(c) that the motion is often helical.  If the ``magnetic'' $\bm{B}$ and ``electrostatic'' $\bm{E}$ fields of Section~\ref{sec:intro} were constant and $E_\perp < B$, then all the motion would be helical.  To see this, apply a Lorentz transformation into a frame moving with velocity $\bm{V} = \bm{E} \times \bm{B}/B^2$. Then $\bm{E}_\perp$ is eliminated, so
$\frac{d \bm{p}_\perp}{d\tau} = {\bf p}_\perp \times \bm{B}$ and $\frac{dp_\parallel}{d\tau} = \gamma E_\parallel$ (using the transformed $\bm{E}$ and $\bm{B}$).  Thus, $\bm{p}_\perp$ rotates at rate $B$ (with respect to proper time) around $\bm{B}$, and $p_\parallel$ changes at rate $E_\parallel$ with respect to coordinate time.  

Note that if $E_\perp > B$ instead, then one can eliminate the component of $\bm{B}$ perpendicular to $\bm{E}$, but we don't pursue that case.

In regions where the ``magnetic'' field is strong we can expect an adiabatic invariant, directly analogous to the magnetic moment for charged particles in strong magnetic fields.  The idea is that if the field seen by a gyrating particle varies by at most a small relative amount $\eps$  during half a gyroperiod then the action integral $\mu = \frac{1}{2\pi} \int \bm{p} \cdot d\bm{r}$ for one gyro-revolution varies by at most order $\eps$ for a number of gyrorevolutions of order $e^{-C/\eps}$ for some $C>0$ (in the analytic case \cite{N}).  The action integral for the unperturbed motion evaluates to 
\begin{equation}
\mu = \frac{p_\perp^2}{B}
\end{equation}
(in plasma physics it is common to divide this by $2m$, where $m$ is the rest mass of the particle, but the above form is simpler in our relativistic context for unit rest mass).
To compute $p_\perp^2$ it is simplest to first compute the field strength 
\begin{equation}
    B = \frac{D}{r^4}\sqrt{R^2+4z^2}
\end{equation}
and parallel momentum
\begin{equation}
    p_\parallel = \tfrac{D}{r^5 B}(3zRp_R + (2z^2-R^2)p_z),
\end{equation}
and use
\begin{equation}
    p_\perp^2 = p^2-p_\parallel^2.
\end{equation}
The total momentum-squared $p^2$ can be computed as
\begin{equation}
    p^2 = p_R^2+p_z^2 + (\tfrac{L_z}{R}-\tfrac{DR}{r^3})^2,
\end{equation}
or 
\begin{equation}
p^2 = (E+\tfrac{1}{r})^2 - 1.
\end{equation}
To complete the computation of $\mu$, $p_\perp^2$ is then divided by $B$. 

To quantify $\eps$ in our problem, the gyroradius $\rho = p_\perp/B = \sqrt{\mu/B}$, the distance travelled by the $\bm{E} \times \bm{B}$ drift in half a gyroperiod is $\gamma \frac{E_\perp}{B}\frac{\pi}{B}$, and the parallel distance travelled during half a gyroperiod is $p_\parallel \pi/B$, so 
$$\eps = \max\left(\frac{2}{L_\perp}\sqrt{\frac{\mu}{B}}, \frac{\pi \gamma E_\perp}{B^2 L_\perp}, \frac{\pi}{L_\parallel} \frac{p_\parallel}{B}\right),$$ 
where $L_\perp$ and $L_\parallel$ are lengthscales for relative variation of $B$ in the two directions.  An order of magnitude for these is $r/3$, and we can take $B \sim D/r^3$ and $|\bm{E}| \sim r^{-2}$.  So the conditions for adiabatic invariance are that $$\mu \ll \frac{D}{r},\ \gamma r^3 \ll {D^2},\ p_\parallel \ll \frac{D}{r^2}.$$  
Note that the first and last imply that $p_\perp^2, p_\parallel^2 \ll \frac{D^2}{r^4}$, so we obtain $\gamma -1 \ll \frac{D}{r^2}$.  Thus the second condition is really just that $r^3 \ll D^2$.

Thus, sufficient conditions for the adiabatic invariance are $r \ll D^{2/3}$, $\mu \ll D^{1/3}$, ${p_\parallel} \ll D^{-1/3}$. 

A consequence of adiabatic invariance of $\mu$ is that the system is close to integrable in the regions where it applies.  Assuming the adiabatic invariance, the motion can be reduced to a 2DoF system for the ``guiding centre'' of the helix, with state-space coordinates being the three components of the guiding centre and its velocity component parallel to the $\bm{B}$-field \cite{Li}.  Rotation symmetry about the $z$-axis then implies 
conservation of $DR^2/r^3$ for the guiding centres (because $B_\phi=0$), which we will write as $r^3 = CR^2$ with $C$ constant.  This is just the equation for staying on a fieldline.  It makes sense because the standard curvature and grad-$B$ drifts across the field are in the $\phi$ direction, which is ignored in the reduction.  
The dynamics of the guiding centre along the fieldline are given by the reduced Hamiltonian 
$$H = \sqrt{1+p_\parallel^2 + \mu B} - \frac{1}{r},$$ 
using arclength $s$ as coordinate. This gives
$$p_\parallel^2 = (E+\tfrac{1}{r})^2 - 1 - \mu B.$$
The field strength $B$ is given by
$$B^2 = \tfrac{D^2}{r^8}(4r^2-3R^2).$$
Parametrise the upper and lower halves of the fieldline by $r \in (0,C]$, $C$ being the value on the equatorial plane, then
$R^2 = \lambda r^3$ with $\lambda = 1/C$.
So
$$p_\parallel^2 = (E+\tfrac{1}{r})^2 - 1 - \tfrac{\mu D}{r^3} \sqrt{4-3\lambda r}.$$
Arclength $s$ is related to $R$ by
$$\left(\frac{ds}{dr}\right)^2 = \frac{1-\tfrac34 \lambda r}{1-\lambda r},$$
which is singular at the extreme ($r=C$) but not a problem.
We see that for $\mu>0$, the motion consists of bouncing along fieldlines between reflection points where 
$$(E+\tfrac{1}{r})^2 = 1 + \tfrac{\mu D} {r^3} \sqrt{4-3\lambda r}.$$

\begin{figure}[htb] 
   \centering
\includegraphics[width=3in]{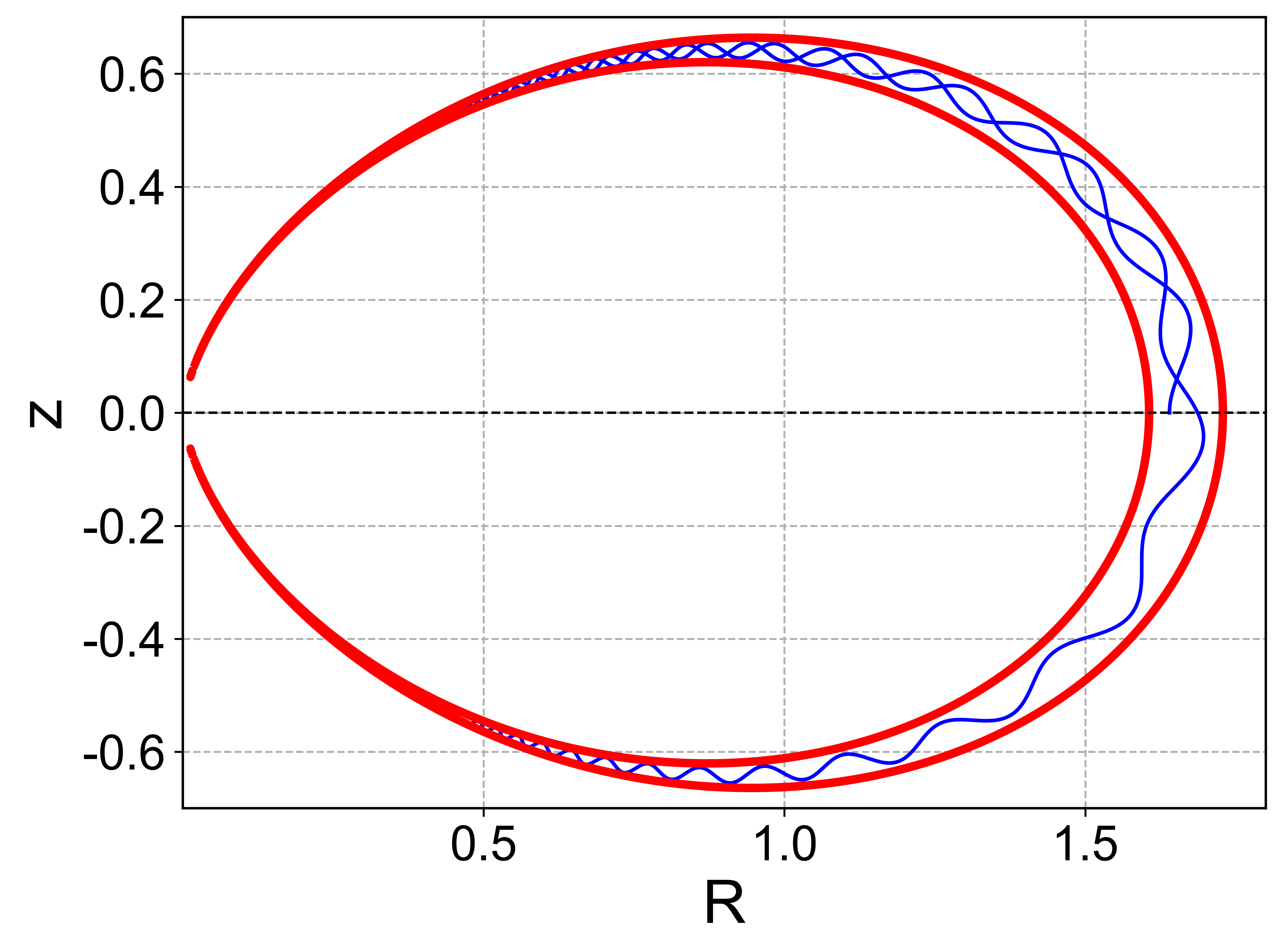}
\includegraphics[width=3in]{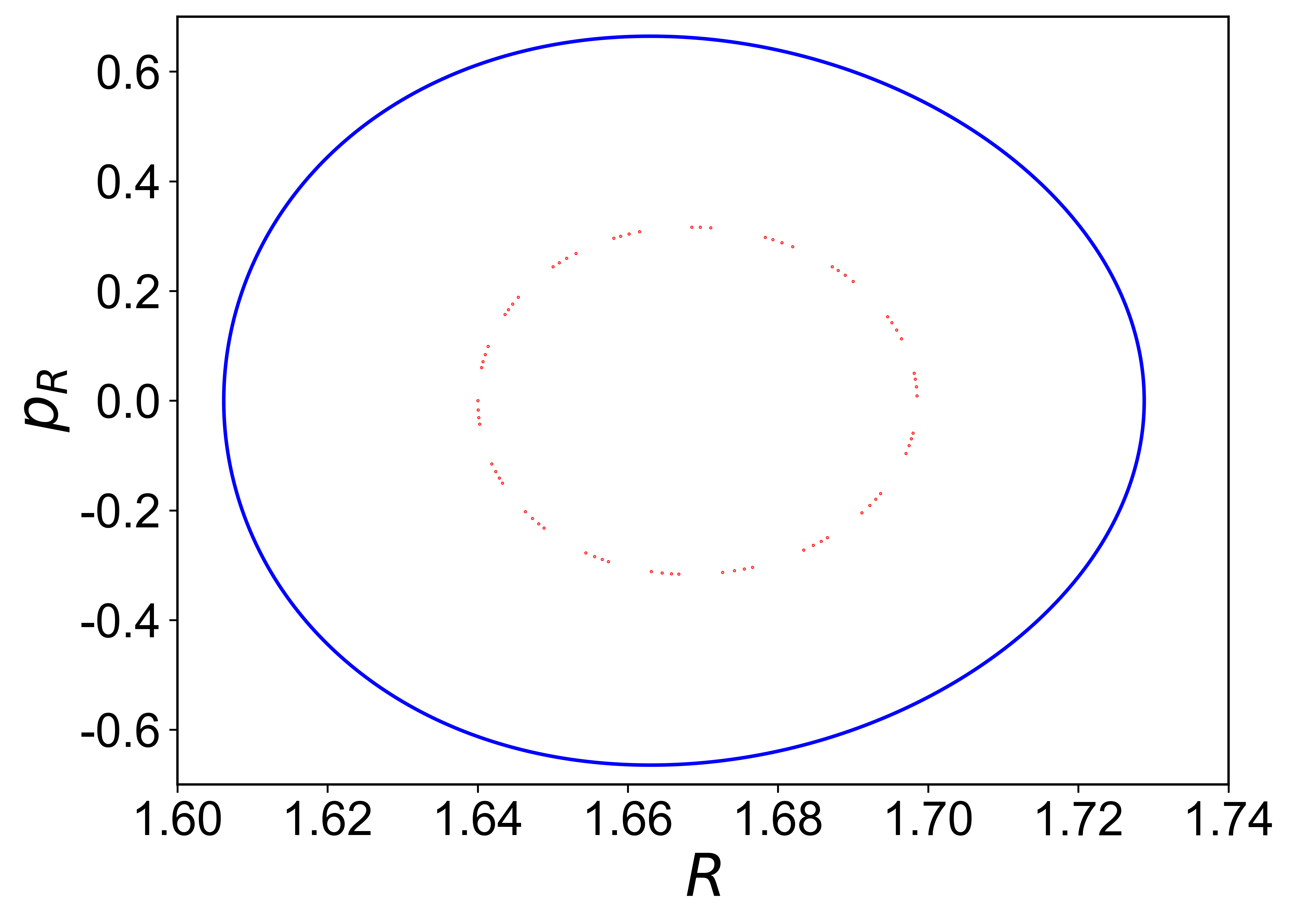}
\caption{(upper) A trajectory for $E=0.6, L_z = 30, D=50$, and (lower) its Poincar\'e plot. 
}
   \label{fig:muconserved}
\end{figure}

So if the adiabatic invariant is well conserved then the return map to a Poincar\'e section $z=0$ for given $E$ and $L_z$ consists of just a rotation by some gyrophase (depending on the gyroradius) about a fixed point, hence ellipses in the $(R,p_R)$-plane.  
Figure~\ref{fig:muconserved} shows an example.
In reality, there is a more accurate adiabatic invariant than $\mu$, which deforms the ellipses into a family of closed curves that includes the boundary, though the adiabatic invariant is probably still not exactly conserved.

\begin{figure}[h!] 
   \centering
\includegraphics[width=3in]{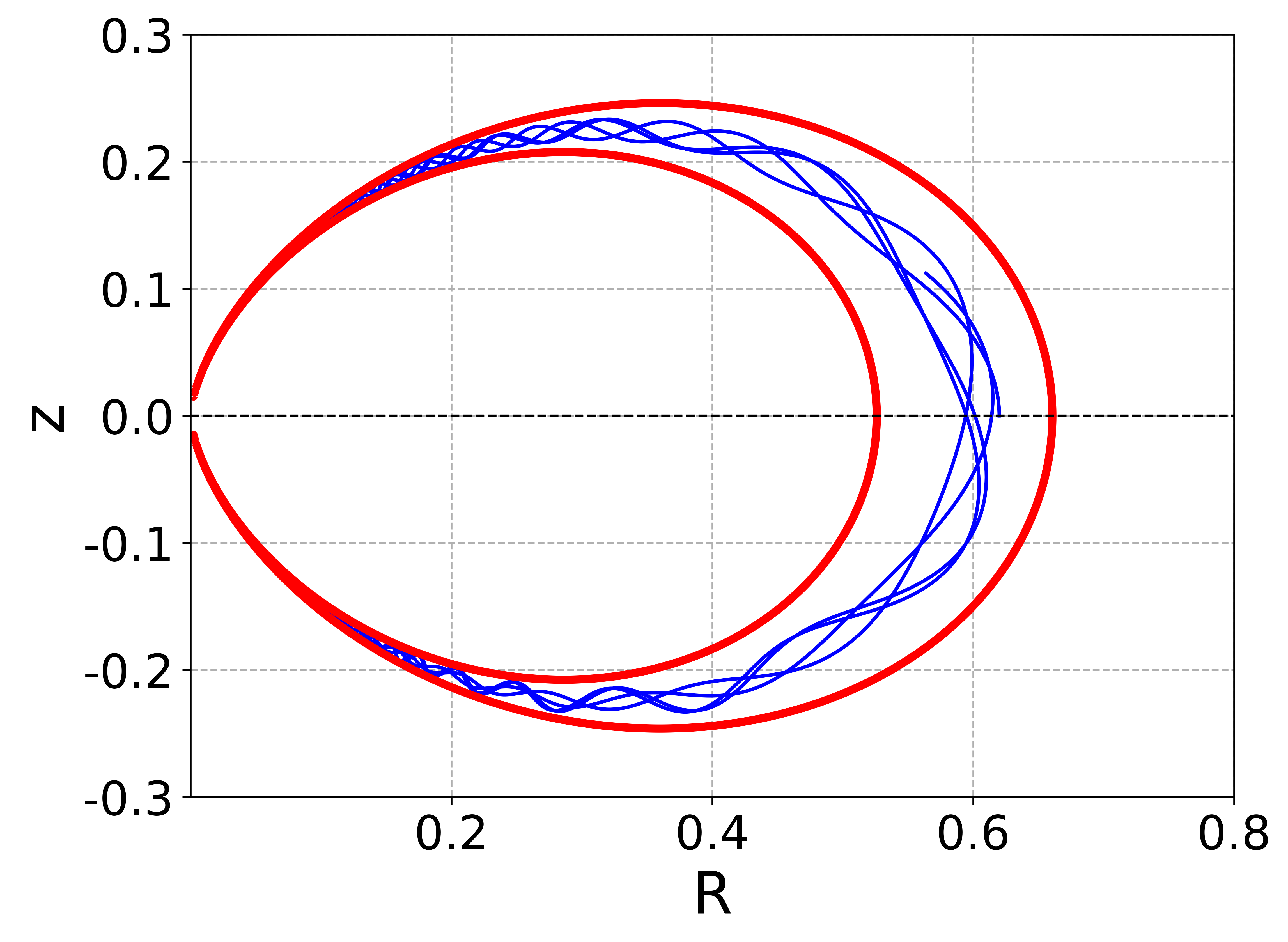}
\includegraphics[width=3in]{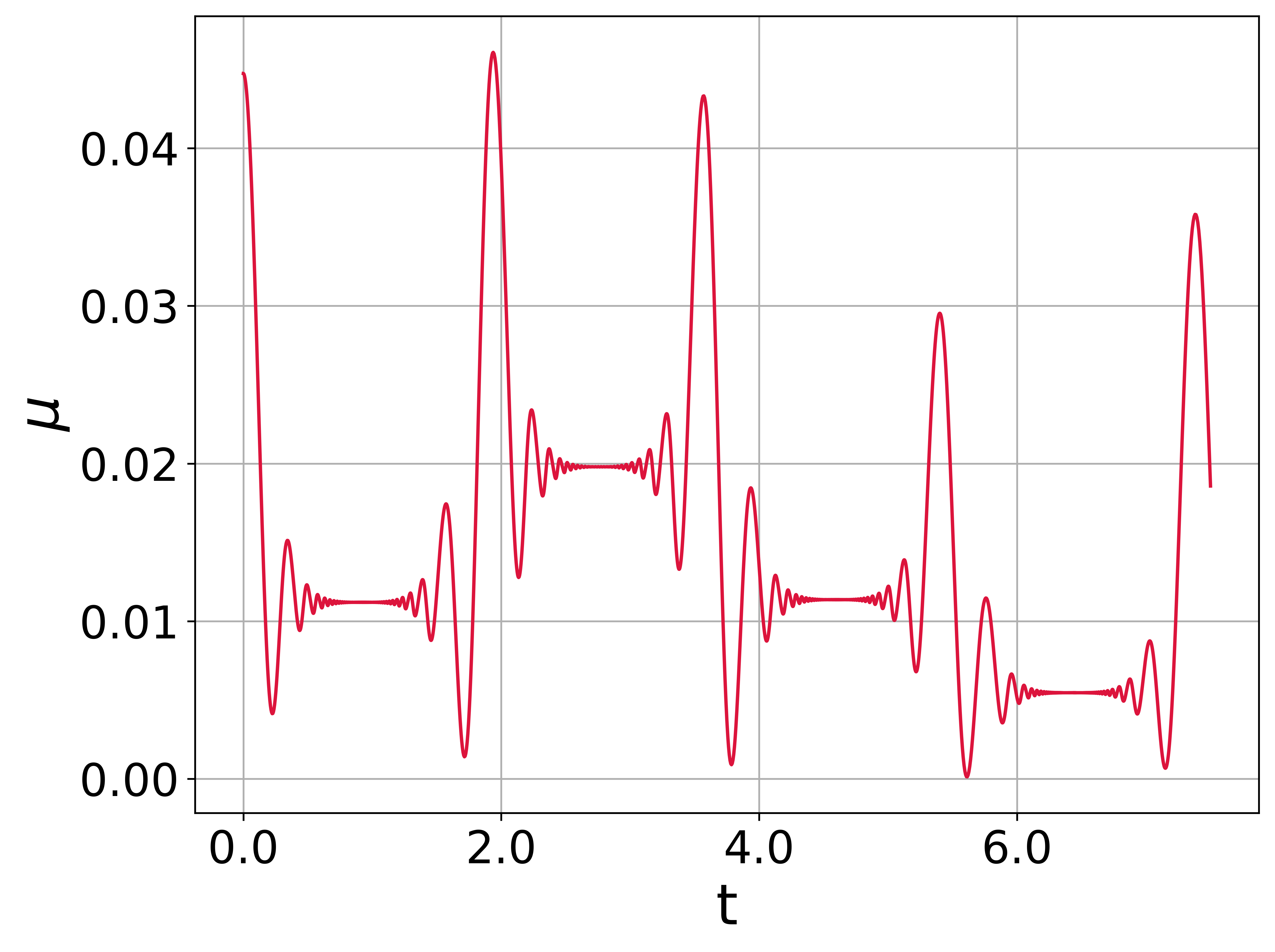}
\caption{(upper) A trajectory for $E=0.9, L_z = 12.5, D=7.3$, exhibiting (lower) fairly good conservation of $\mu$ while in the horns and substantial changes in between. 
}
   \label{fig:adiab}
\end{figure}

In practice, what we found more often (particularly when $L_z>0$, for which the Hill components have horns going to the origin) is that the adiabatic invariant is well conserved while the particle is in the horns but not in the passage at relatively large radius between the horns (it moves too far along the field in one gyroperiod).  An example is shown in Figure~\ref{fig:adiab}.

Such pictures suggest an alternative approach to understanding the near-perfect chaos.  Chaos in a Hamiltonian system can result from possession of an adiabatic invariant in parts of the phase space, but with pseudo-random jumps in the adiabatic invariant in between visits to these parts.  The failure of the adiabatic invariant in most examples studied so far is due to separatrix crossing (where the normally rapid frequency goes to zero) \cite{N+}.  In our system we do not see separatrix crossing, but it is clear from pictures like Figure~\ref{fig:adiab} that the motion sometimes consists of episodes in the cusps of the Hill's region, where the adiabatic invariant is well conserved, separated by passage across $z=0$ via relatively large radius, during which some change to the adiabatic invariant occurs.  Perhaps one can make a probabilistic analysis of the jumps in adiabatic invariant, as has been done for cases of separatrix crossing, and deduce stochastic behaviour.  This approach would be limited to systems like ours with an adiabatic invariant, so not as general as the approach of the previous section.  Yet it could be simpler in cases like ours when it applies.

\section{Rotation curves of galaxies}
\label{sec:rotcurve}

Aubry's investigation was motivated by the desire to explain the rotation curves of galaxies without invoking dark matter (for another approach, which prompted Aubry's investigations, see ref.~\onlinecite{Ro}).  Here we see what our investigations can say about the issue.

The rotation curve of a galaxy is the graph of the inferred average of $v_\phi$ as a function of $R$ in the equatorial plane.  For many galaxies it is observed to settle to a constant around $10^{-3}$ (in units of $c$) as $R \to \infty$.  The observations are based on Doppler shift of $H_\alpha$ lines from the gas in the disk.

The co-rotating equatorial circular orbits are the simplest theoretical probe for this question, but they have $v_\phi \propto R^{-1/2}$ as $R \to \infty$, so they do not fit the observations.  This is the pure Newtonian prediction but we see from (\ref{eq:circeq}) that it is true also for Aubry's galaxy. 

The standard explanation is that in addition to the visible matter in a galaxy, there is a halo of ``dark matter'' whose gravitational effect makes $v_\phi$ for the circular orbits go to a positive limit as $R \to \infty$.  
For one analysis, see ref.~\onlinecite{KT}.

But here is another approach, avoiding the appeal to dark matter.  We have seen that for orbits of Aubry's system,
\begin{eqnarray}
\gamma &=& E+\tfrac{1}{r} \\
p_\phi &=& \tfrac{L_z}{R}-\tfrac{DR}{r^3}.
\end{eqnarray}
To fit with a non-zero limit for tangential velocity as $R \to \infty$, we need $p_\phi$ to tend to some  $p_\infty>0$ (one could consider $p_\infty < 0$ instead, but that is unlikely).  So 
\begin{equation}
L_z = R p_\infty + \tfrac{DR^2}{r^3}
\end{equation}
must grow with $R>\sqrt{D/p_\infty}$ (we take $r \sim R$ in the disk).
Furthermore, 
\begin{equation}
E = \sqrt{1+ p_\infty^2 + p_R^2+p_z^2} - \tfrac{1}{r}
\end{equation}
must decrease unless $p_R^2+p_z^2$ makes a compensating increase.
Different $L_z$ for different $R$ requires different orbits, hence circular, and there are none with $\gamma$ bounded away from 1 as $R \to \infty$.
So instead perhaps there is exchange of $L_z$ and of $E$ between particles, leading to a distribution that is concentrated around $L_z = Rp_\infty + \frac{D}{R}$ and $E = \gamma_\infty - \frac{1}{R}$? This distribution would necessarily have a nett outward flux of particles for $R > \frac{2}{p_\infty^2}$ because then $E>1$ and hence $\ddot{R}>0$, but that is not necessarily a problem.  Why shouldn't a galaxy be losing gas?

It might also be that matter goes in or out along the ``horns'' of the Hill's regions with $L_z>0$.  Rourke \cite{Ro} proposed outflow along the rotation axis with gradual inflow in the equatorial disk, but perhaps it is the other way round.

To develop this idea requires some kinetic theory, so we leave it for future work.  

Another ingredient to consider is that there could be transfer between orbital angular momentum (our $L_z$) and spin (angular momentum of a particle about its centre of mass), which we have neglected.

Perhaps neither of these effects depends significantly on $D$, so Aubry's model might be more sophisticated than necessary to explain the rotation curves of galaxies.


\section{More Comments}

In general relativity, it has long been recognised that rotating matter has ``frame-dragging'' effects \cite{MTW}.  The gravitomagnetic force used here is a simple manifestation of frame-dragging.  To our minds, it demystifies frame-dragging because it shows frame-dragging to correspond to the familiar effects of a magnetic field.

Rourke's proposal \cite{Ro} to explain rotation curves of galaxies (and more) involves modifying the standard Lense-Thirring frame-dragging effect to one that decays like only $1/r$ with distance from a rotating mass.  He justifies it by Mach's principle, but it is not clear to us.

One direction to explore is the possible connection of the magnetic gravity term with Thomas precession \cite{BM}, which is usually considered at the atomic scale but applies more broadly.

For non-relativistic motion in Aubry's galaxy, one could study the non-relativistic version of (\ref{eq:Ham}).  Then $c$ becomes irrelevant and one can scale $D$ to $1$.  A question that Aubry has posed is what is the set of orbits that go to neither infinity nor zero.  This could predict the shape of an axisymmetric galaxy.  It is an interesting direction for future research.

Another question Aubry posed is what are the effects of the gravitomagnetic field on light rays?  In our model, however, the effect of putting rest mass to zero is to remove all effects of gravity and so to just produce constant velocity trajectories at the speed of light.

Another direction to explore is to take gravitational force proportional to relativistic mass instead of rest mass.  This would in particular, produce an effect on light rays.  We didn't find a Hamiltonian formulation of this case, except when $D=0$.  That is not necessarily an obstacle to studying it, but means it would require other methods.  Nonetheless, for any $D$ an ``energy'' is conserved, namely $H=\log\gamma -\tfrac{1}{r}$. Equivalently, $\gamma\, \exp-\tfrac{1}{r}$ is conserved, which supports a point of view taken by Aubry\cite{A}, namely that the effect of a static potential can be replaced by taking the rest mass $m$ of a particle to decrease as it goes down the potential. The total energy of the particle then being $\gamma m$, we obtain the formula $m \propto \exp-\tfrac{1}{r}$.
Note also that standard volume is preserved if one uses a new time $s$ with $ds = \gamma\, dt$.  Thus, the system is close to Hamiltonian, but we did not yet identify an appropriate symplectic form, nor even angular momentum constant.  There is a vague possibility that one should view conservation of $\gamma\, \exp-\tfrac{1}{r}$ as a non-holonomic velocity-constraint, so rotation-symmetry might not produce a conservation law \cite{Bl}.

Finally, it would be natural to consider the effects of breaking axisymmetry.  Spiral galaxies have clearly done that.  Then $L_z$ conservation would no longer be perfect.

\section{Conclusion}

We have studied the motion of a test mass in the gravitomagnetic field proposed by Aubry as a simple model of a galaxy, taking force proportional to rest mass.  
The system can equivalently be viewed as motion of a relativistic charged particle in the field of a co-located electrostatic monopole and magnetic dipole.
We find a mix of regular and chaotic motion.  The dynamics has various forms, depending on the energy and angular momentum constants.  An interesting regime of almost perfect chaos is found, whose analysis merits more investigation.  Further directions for analysis of possible consequences for galaxies are proposed.

\begin{acknowledgments}
RSM is most grateful to Serge Aubry for stimulation over many years.  This began with Aubry's fascinating talk on minimisers of Frenkel-Kontorova models at Los Alamos in 1982.  Notable other highlights were his explaining anti-integrable limits to me one Sunday morning in Minnesota in 1990, and setting me the challenge of explaining discrete breathers in 1993 during a visit to Saclay.
Most recently he communicated to me his ideas on rotation curves of galaxies, which led to the present paper.

We are grateful to Shibabrat Naik for making professional versions of some of our hand-drawn figures and to Vassili Gelfreich for useful suggestions.

This work was supported by a grant from the Simons Foundation (601970, RSM).

\end{acknowledgments}

\section*{Data Availability Statement}

The data that support the findings of this study are openly available in GitHub at https://github.com/megha-burri/Motion-in-Aubry-s-galaxy.git and archived on Zenodo at https://doi.org/10.5281/zenodo.17117918.

\end{document}